\documentclass[
preprint, nofootinbib,amsmath,amssymb,aps
]{revtex4-2}

\usepackage{bm}
\usepackage{physics, amsmath, amsfonts, siunitx, tensor, paralist, comment, setspace}
\usepackage[font=small,labelfont=bf,justification=justified, singlelinecheck = false]{subcaption}
\usepackage[font=small,labelfont=bf,justification=raggedright, singlelinecheck = false]{caption}
\usepackage{graphicx, color}
\usepackage{hyperref}
\usepackage[mathlines]{lineno}
\hypersetup{%
bookmarksnumbered=true,%
bookmarksopen=true,%
colorlinks=true,%
linkcolor=blue,
citecolor=red,
}

\allowdisplaybreaks[4]

\newcommand\nn{\nonumber \\}
\newcommand\e{\mathrm{e}}

\newcommand{\figref}[1]{FIG.~\ref{#1}}
\newcommand{\tabref}[1]{TABLE~\ref{#1}}
\newcommand{\secref}[1]{Sec.~\ref{#1}}
\newcommand{\appref}[1]{Appendix.~\ref{#1}}

\begin{document}
\count\footins = 1000

\title{Revisiting compact star in \texorpdfstring{$F(R)$}{F(R)} gravity: \\
Roles of chameleon potential and energy conditions}

\author{Kota Numajiri}%
\email{numajiri.kota.m3@s.mail.nagoya-u.ac.jp}
\affiliation{
Department of Physics, Nagoya University, Nagoya 464-8602, Japan
}

\author{Yong--Xiang Cui}
\email{cyx2021112305@mails.ccnu.edu.cn}
\affiliation{
Institute of Astrophysics, Central China Normal University, Wuhan 430079, China
}

\author{Taishi Katsuragawa}
\email{taishi@ccnu.edu.cn}
\affiliation{
Institute of Astrophysics, Central China Normal University, Wuhan 430079, China
}

\author{Shin'ichi~Nojiri}
\email{nojiri@gravity.phys.nagoya-u.ac.jp}
\affiliation{
Department of Physics, Nagoya University, Nagoya 464-8602, Japan
}
\affiliation{
Kobayashi-Maskawa Institute for the Origin of Particles and the
Universe, Nagoya University, Nagoya 464-8602, Japan
}

\begin{abstract}
We reexamine the static and spherical symmetric compact star configuration in the $R^2$ model of the $F(R)$ gravity theory.
With asymptotic solutions for the additional scalar degrees of freedom,
we refine analysis on the external geometry and settle the scalar-hair problem argued in previous works.
Performing the numerical integration of the modified Tolman-Oppenheimer-Volkoff equations as a two-boundaries-value problem, we further discuss the scalar-field distribution inside the compact stars and its influence on the mass-radius relation. 
We show that the chameleon potential plays an essential role in determining the scalar field inside the star. 
The scalar field often behaves as a quintessential field that effectively decreases the mass of compact stars with lower central energy density.
\end{abstract}

\maketitle
\newpage
\tableofcontents

\section{Introduction}

In spite of the empirical successes of General relativity (GR), it is known that there are several problems in a wide range of energy scales, from the dark sector problems to the quantization of the gravitational field. 
To tackle these problems within the gravity sector, the modification of GR, which we call the modified gravity theories, has been considered. 
These theories usually introduce new degrees of freedom (DOF) for the gravitational field, and additional DOFs enable us to solve the problems in GR (see \cite{Nojiri:2010wj, Nojiri:2017ncd, Arai:2022ilw} for review).
The modified gravity theories (and GR) are also expected to be the effective field theory for the quantum gravity theory, and we may obtain hints about the most fundamental theory from the study on modified gravity theories.

One of the most popular classes in modified gravity theories is the $F(R)$ gravity theory (see~\cite{Sotiriou:2008rp, DeFelice:2010aj, Faraoni:2010pgm} for review). 
The $F(R)$ gravity theory is obtained by replacing the Ricci scalar term $R$ in the Einstein-Hilbert action with a function of $R$, $F(R)$. 
This modification introduces an additional scalar DOF, so-called scalaron, to the gravitational field,
and such a new scalar field can drive the accelerated expansion of the Universe in the early and late-time epoch~\cite{Capozziello:2002rd, Nojiri:2007as, Cognola:2007zu, Tsujikawa:2007xu}.
Several works have suggested that the scalaron can also serve as the dark matter and affect the structure formations in the Universe~\cite{Katsuragawa:2016yir, Katsuragawa:2017wge, Yadav:2018llv, Shtanov:2021uif, Shtanov:2021uif, KumarSharma:2022qdf, Shtanov:2022xew}.
As one of the $F(R)$ models, this paper focuses on the $R^2$ gravity model that includes an additional curvature-squared term to the Einstein-Hilbert action. 
This model is known to realize the inflation dynamics by the gravity sector, which is so-called the Starobinsky inflation model \cite{Starobinsky:1980te}, as we tune a coefficient of the $R^2$ term.

Recently, the compact star configuration has been one of the main interests in the modified gravity theories.
In addition to the study on black hole solutions in $F(R)$ gravity~\cite{delaCruz-Dombriz:2009pzc, Cembranos:2011sr, Sheykhi:2012zz, Tang:2019qiy, Khodadi:2020cht, Khodadi:2022xtl}, 
static and spherically symmetric solutions for neutron stars in the $R^2$ gravity have been investigated in both the perturbative approach \cite{Arapoglu:2010rz} and non-perturbative approach \cite{Ganguly:2013taa, Yazadjiev:2014cza, Capozziello:2015yza, Resco:2016upv, Astashenok:2017dpo, Feng:2017hje, Astashenok:2018iav, Nava-Callejas:2022pip}. 
Similar considerations with the torsion and axion field were performed~\cite{Feola:2019zqg, Astashenok:2020isy}, respectively. 
In Refs.~\cite{Yazadjiev:2014cza, Capozziello:2015yza, Resco:2016upv, Astashenok:2017dpo, Astashenok:2018iav, Feola:2019zqg}, 
the mass-radius relation of the neutron stars was calculated.
These previous works generally showed that the relation curve is distorted, and the maximum mass increases regardless of the equation of state (EOS) as the modification of the theory becomes dominant. 
This result may explain the observed existence of massive neutron stars with $\sim 2.0 \, M_\odot$ \cite{Demorest:2010bx, Antoniadis:2013pzd} from the view of the gravity theory. 
The quark stars and the white dwarfs in the $R^2$ gravity were also considered~\cite{Astashenok:2014dja, Astashenok:2022kfj}.

We highlight some remaining issues in those previous works. 
One is the external spacetime around the compact star system. 
In the general $F(R)$ gravity, the static and spherical vacuum solution is not always the Schwarzschild spacetime due to the absence of alternatives to Birkhoff-Jebsen's theorem in GR. 
In the $R^2$ gravity theory, the typical behaviors for the asymptotically flat solution have not been specified. 
For instance, in Refs.~\cite{Yazadjiev:2014cza, Astashenok:2017dpo, Astashenok:2018iav}, they derived the outer solution numerically and revealed that the neutron stars in $R^2$ gravity have monotonically decaying scalar hair under the asymptotic flatness condition. 
On the other hand, in Refs.~\cite{Resco:2016upv, Feola:2019zqg}, it was indicated that scalar hairs damp with oscillation for the stable branch against radial perturbations. 
Moreover, Ref.~\cite{Ganguly:2013taa} claimed that the external solution must coincide with the Schwarzschild solution exactly and the neutron stars have no scalar hair due to the no-hair theorem proved for the static and spherical black hole system~\cite{Whitt:1984pd, Mignemi:1991wa, Nzioki:2009av}. 
In this way, even qualitative expectation for outer solutions has yet to be confirmed consistently. 
Another is the behavior of the scalaron field (or equivalently the curvature) inside the star. 
Most previous works have been interested in the mass and radius of compact stars and have not paid much attention to the inner profile of the scalaron field and its cause. 
This viewpoint is necessary for understanding the inner structure and interactions of the compact stars with modified gravity theories which have additional DOFs for the gravity sector.

In this paper, we reexamine the static and spherically symmetric star configuration in the $R^2$ gravity.
We utilize the asymptotic behavior of the scalaron field to tackle the controversial discussion on external geometry and to improve the analysis of internal geometry and scalaron field distribution.
We pay special attention to the fact the scalaron field is determined by the effective potential called the chameleon potential~\cite{Khoury:2003rn, Brax:2008hh, Burrage:2017qrf}. 
This potential has been often employed to suppress the scalaron field propagation and to reproduce the observed validity of GR~\cite{Capozziello:2007eu, Moretti:2019yhs, Katsuragawa:2019uto}, which is one of the so-called screening mechanisms.
As we will see later, $R^2$ gravity does not possess the screening mechanism. However, the chameleon potential is helpful in analyzing the scalaron field (i.e., the effects of modifying the gravitational theory) regardless of whether the screening mechanism works. 
Similar investigations of the scalaron field on the compact star configuration under other types of the $F(R)$ models can be found~\cite{Kobayashi:2008tq, Babichev:2009fi}.

We apply the above viewpoints of the chameleon potential and the gravity configuration to the compact star system. 
Firstly, we specify the physically realizable asymptotic geometry by probing the asymptotic solution for the scalar DOF. 
We demonstrate that the analytical approach to the scalaron field distribution gives insights into the asymptotic geometry and the proper integration method for numerical calculations. 
Secondly, based on the above consideration, we numerically solve the modified Tolman–Oppenheimer–Volkoff (TOV) equations in the case of neutron stars. 
We show how the modification of the gravity theory influences the structure, geometry, and observables, such as the mass-radius relation.
Focusing on the chameleon potential and the energy conditions of the scalaron field, we also argue that these influences are clearly correlated to the role of the scalaron field inside the compact star.

The remaining part of this paper is organized as follows: 
We briefly review the $F(R)$ gravity, the $R^2$ model, and the chameleon potential in \secref{sec:basics}, and then we discuss the desired static and spherical symmetric compact star configuration and derive the modified TOV equations in \secref{sec:system}. 
The setting of the actual numerical calculation and the results with discussions are given in \secref{sec:results}. 
The \secref{sec:conclusion} is dedicated to the conclusion. 
In this work, we use $c=G=1$ unit basically, and other notations follow Ref.~\cite{carroll_2019}.


\section{Basics and Chameleon Mechanism of \texorpdfstring{$R^2$}{R^2} Gravity \label{sec:basics}}

The action of the gravitational field in the $R^2$ gravity model~\cite{Starobinsky:1980te} is defined as
\begin{align}
    S_G = \frac{1}{2\kappa^2} \int d^{4}x \sqrt{-g} \; F(R) \, ,
    \label{eq:action_Jordan}
\end{align}
where
\begin{align}
    F(R) = R + \alpha R^2 \, ,
    \label{eq:R^2}
\end{align}
and $\kappa^2 = 8\pi G$ is the gravitational coupling constant. 
The limit $\alpha \rightarrow 0$ reduces to the usual Einstein-Hilbert action. 
The frame with the gravitational field action in this form is called the Jordan frame~\footnote{We can transform it into the form of Einstein gravity with a canonical scalar field using the scale transformation. 
This frame is called the Einstein frame in contrast. 
See~\appref{sec:EinsteinFrame}.}. 

Performing the variation of the gravitational action~\eqref{eq:action_Jordan} and matter-sector action with respect to the metric $g_{\mu\nu}$, we obtain the field equation:
\begin{align}
    &F_R (R)\, R_{\mu\nu} - \frac{1}{2} F(R) \, g_{\mu\nu} 
    + \qty(g_{\mu\nu}\Box - \nabla_{\mu}\nabla_{\nu}) \, F_R (R)
    = \kappa^2 T_{\mu\nu} \, ,
    \label{eq:EOM_Jordan}
\end{align}
where $F_R = dF/dR$.
The energy-momentum tensor is defined by the matter action $S_M$ as
\begin{align}
    T_{\mu\nu} = \frac{-2}{\sqrt{-g}}
    \fdv{S_M}{g^{\mu\nu}} \,.
\end{align}
Taking trace of Eq.~\eqref{eq:EOM_Jordan}, we find
\begin{align}
    \Box F_R (R)
    = \frac{1}{3} \qty[2 F(R) - R F_R (R) + \kappa^2 T] \, ,
    \label{eq:trEOM_Jordan}
\end{align}
where $T$ represents the trace of energy-momentum tensor $T=T\indices{^\mu_\mu}$. 

It should be noted that the trace equation~\eqref{eq:trEOM_Jordan} takes nontrivial form even for the vacuum $T=0$, as opposed to $R=0$ in the GR case. 
Thus Birkhoff-Jebsen's theorem~\cite{birkhoff1923relativity, jebsen2005general}, which states the Schwarzschild solution is a unique solution for a spherically symmetric vacuum system in GR, is generally absent in the $F(R)$ gravity. 
As a result, vacuum, static, and spherically symmetric solutions with nontrivial curvature distribution (i.e., scalarized solutions) are allowed \cite{Nashed:2021lzq}. 
This fact stems from the additional scalar DOF, which we will mention later, and it is one of the crucial points in this paper. 

In the $R^2$ gravity~\eqref{eq:R^2}, the field equations Eqs.~\eqref{eq:EOM_Jordan} and \eqref{eq:trEOM_Jordan} reduce to
\begin{align}
    R_{\mu\nu} - \frac{1}{2} R \, g_{\mu\nu}
    + \alpha R \qty(2 R_{\mu\nu} - \frac{1}{2}R\, g_{\mu\nu})
    + 2 \alpha \, \qty(g_{\mu\nu}\Box - \nabla_{\mu}\nabla_{\nu}) \, R
    = \kappa^2 T_{\mu\nu} \, ,
\end{align}
and 
\begin{align}
    2 \alpha \Box R 
    = \frac{1}{3} \qty[ R + \kappa^2 T] \, ,
    \label{eq:trEOM_Jordan_R^2}
\end{align}
respectively. 
The $\alpha \rightarrow 0$ limit recovers the Einstein equation and the curvature--energy-momentum relation in GR again.

It is well known that the $F(R)$ gravity can be rewritten in the form of the scalar-tensor theory. 
We introduce the scalar field, called scalaron, $\Phi \equiv F_R (R)$, which one can solve with respect to $R$ as $R=\Tilde{R}(\Phi)$. 
Then the action~\eqref{eq:action_Jordan} is rewritten as
\begin{align}
\begin{split}
    &S_G = \frac{1}{2\kappa^2} \int d^{4}x \sqrt{-g} \,
    \qty[ \Phi R - V(\Phi)]\, , \\
    &V(\Phi) = \Tilde{R} (\Phi) \Phi - F\qty(\Tilde{R}(\Phi))\, .
\end{split}
    \label{eq:action_BD}
\end{align}
The field equations for the metric $g_{\mu\nu}$ and scalaron $\Phi$ are
\begin{align}
    &R_{\mu\nu} - \frac{1}{2} g_{\mu\nu} R
    =\frac{\kappa^{2}}{\Phi} 
    \qty(T_{\mu \nu} + T^{(\Phi)}_{\mu \nu} ), 
    \label{eq:EOM_metric_BD}\\
    &\Box \Phi 
    = \frac{1}{3} \qty[\Phi V'(\Phi) - 2V(\Phi) + \kappa^2 T] \, , 
    \label{eq:EOM_scalar_BD}
\end{align}
where $T_{\mu\nu}^{(\Phi)}$ is the effective energy-momentum of the scalaron field:
\begin{align}
    T^{(\Phi)}_{\mu \nu} = 
        \frac{1}{\kappa^2}
        \qty[
            \nabla_{\mu} \nabla_{\nu} \Phi-g_{\mu \nu} \square \Phi
            -\frac{1}{2} g_{\mu \nu} V(\Phi)
        ]\, .
\end{align}
Therefore the gravitational field in the $F(R)$ gravity act as the Einstein gravity with a non-minimally coupled scalar field, which has $2+1$ DOF.

For the $R^2$ model~\eqref{eq:R^2}, the scalaron field $\Phi$ is defined as the linear form of the curvature: 
\begin{align}
    \Phi \equiv F_R (R)
    = 1 + 2 \alpha R\, .
    \label{eq:scalar_BD_R2}
\end{align}
We can use the curvature as the independent field for the scalar DOF in this gravity model, instead of using only metric $g_{\mu\nu}$ for both tensorial and scalar DOF.
The equation for the scalaron field and its energy-momentum tensor are given as
\begin{align}
    \Box \Phi 
    = \frac{1}{6\alpha} \qty[\Phi -1 + 2 \alpha \kappa^2 T] \, , 
    \label{eq:EOM_scalar_BD_R^2}
\end{align}
and
\begin{align}
\begin{split}
    &T^{(\Phi)}_{\mu \nu} = 
        \frac{1}{\kappa^2}
        \qty[
            \nabla_{\mu} \nabla_{\nu} \Phi-g_{\mu \nu} \square \Phi
            -\frac{1}{2} g_{\mu \nu} V(\Phi)
        ]\, ,
    \\
    &V(\Phi) = \frac{1}{4\alpha} \qty(\Phi-1)^2\, ,
\end{split}
        \label{eq:EMtensor_scalar}
\end{align}
respectively. 
We can notice that Eq.~\eqref{eq:EOM_scalar_BD_R^2} is merely the rewording of Eq.~\eqref{eq:trEOM_Jordan_R^2}, and thus we can use Eq.~\eqref{eq:trEOM_Jordan_R^2} as the field equation for curvature $R$. 

Now we should remark on the value of $\Phi$. 
From the condition of no anti-gravity interaction, Eq.~\eqref{eq:EOM_metric_BD} suggests that the value of the scalaron field $\Phi$ must be positive. 
This positivity generally restricts the modification parameter ($\alpha$ in the $R^2$ gravity) or the curvature value $R$. 
In the $R^2$ gravity which we are dealing with, the absence of anti-gravity gives a bound on the possible values of curvature ($R > -1/2\alpha$ for positive $\alpha$ and $R < 1/2|\alpha|$ for negative $\alpha$). 

\begin{figure}[tb]%
  \begin{minipage}[t]{0.5\linewidth}%
    \centering%
    \includegraphics[keepaspectratio, width=0.9\linewidth]{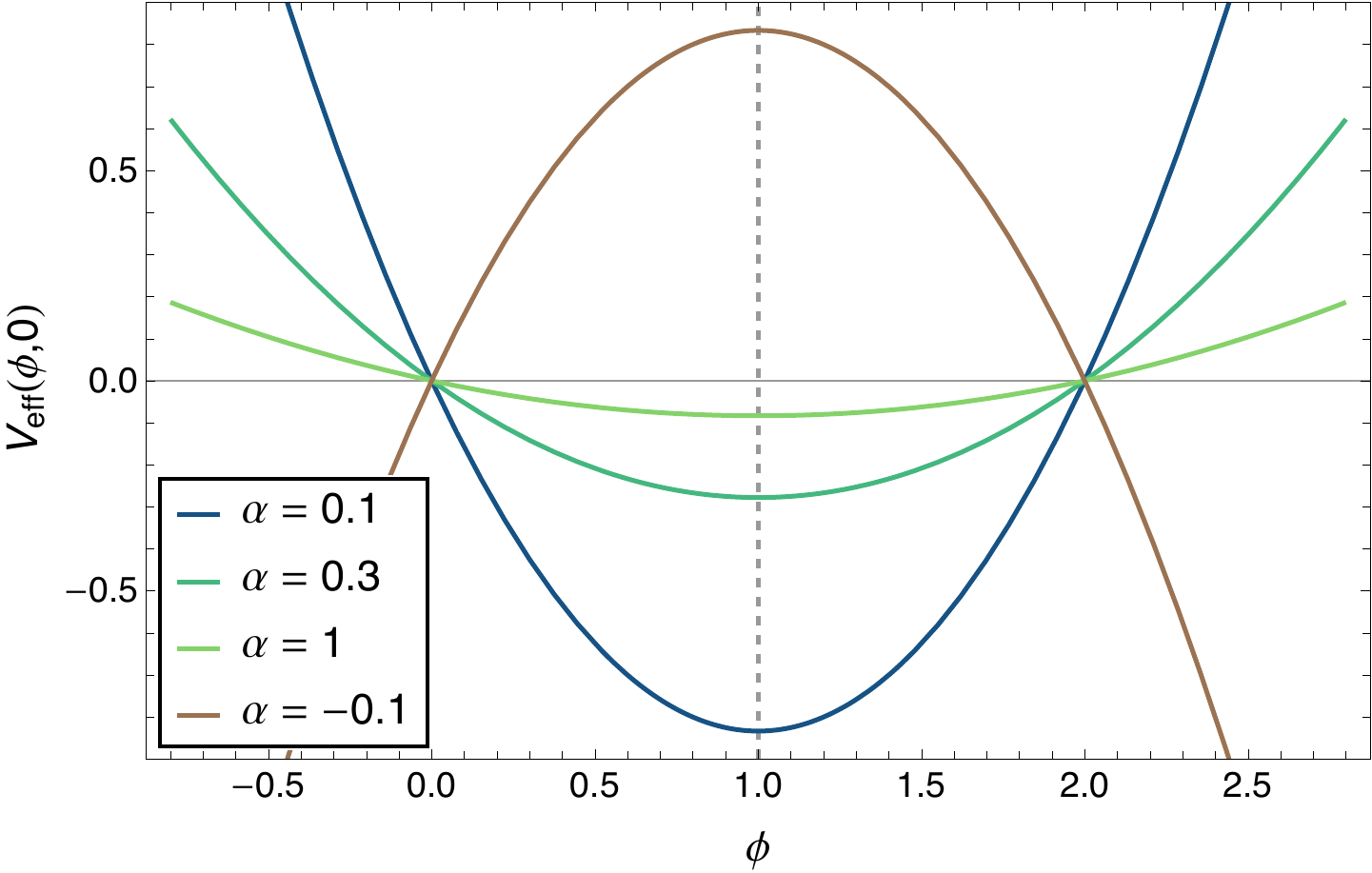}%
    \subcaption{Varying $\alpha$ with $T=0$.}%
    \label{fig:chame_pot_JO_a}%
  \end{minipage}%
  \begin{minipage}[t]{0.5\linewidth}%
    \centering%
    \includegraphics[keepaspectratio, width=0.9\linewidth]{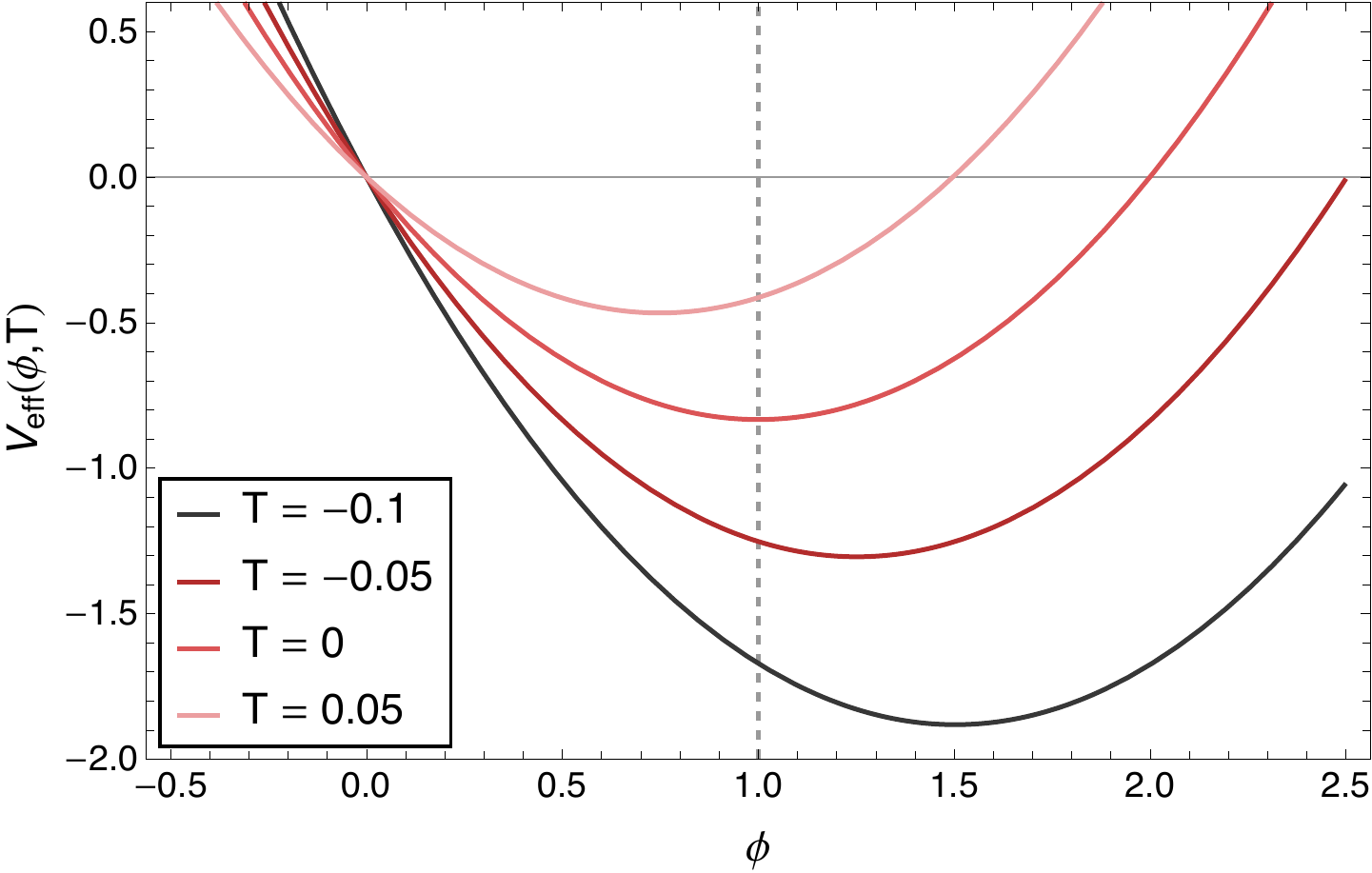}%
    \subcaption{Varying $T$ with $\alpha=0.1$.}%
    \label{fig:chame_pot_JO_T}%
  \end{minipage}%
  \caption{
The chameleon potential~\eqref{eq:cham_pot_R^2_Jordan} in the Jordan frame under the $R^2$ gravity~\eqref{eq:R^2}. 
All quantities are dimensionless by half of the Schwarzschild radius for the solar mass $r_g = GM_{\odot}$. 
The potential becomes a convex function for $\alpha>0$ or a concave one for $\alpha<0$, and negative $\alpha$ causes instability for the scalaron field. The potential shape becomes steeper as $|\alpha|$ decreases. 
One can see from (b) that the changes of $T$ lead to translations of the bottom of potential. 
}
\label{fig:chame_pot_JO}
\end{figure}%

The right-hand side of the scalaron field 
equation~\eqref{eq:EOM_scalar_BD} can be recognized as the force term which stems from the effective potential $V_{\mathrm{eff}} (\Phi, T)$:
\begin{align}
    \Box \Phi = \frac{1}{3} \qty[\Phi V'(\Phi) - 2V(\Phi) + \kappa^2 T]
    \equiv \pdv{V_{\mathrm{eff}}}{\Phi} \qty(\Phi, T) 
    \, .
    \label{eq:eom_scalar_cham}
\end{align}
This energy-momentum-dependent potential is called the chameleon potential \cite{Khoury:2003rn}, 
where the scalaron field is often called the chameleon field, 
and we define the effective mass of the scalaron field by expanding the potential around its minimum $\Phi=\Phi_{\min}$ where $\partial V_{\mathrm{eff}} (\Phi_{\min}) / \partial \Phi = 0$ holds:
\begin{align}
    m_\Phi^2 =\eval{\pdv[2]{V_{\mathrm{eff}}}{\Phi}}_{\Phi=\Phi_{\min}}
    = \frac{1}{3} \qty[
        \frac{F_R \qty(R(\Phi_{\min}))}{F_{RR} \qty(R(\Phi_{\min}))}-R(\Phi_{\min})
    ]\, .
    \label{eq:cham_mass_Jordan}
\end{align}
As Eq.~\eqref{eq:cham_mass_Jordan} shows, the chameleon potential and scalaron mass in general depend on $T$ and control the propagation of the scalaron field. 
The $T$-dependence generates so-called the screening mechanism in some classes of the $F(R)$ gravity~\cite{Khoury:2003rn}.
For high $|T|$ regions, such as the atmosphere of the star~\cite{Katsuragawa:2019uto}, 
the scalaron mass can be very large, and the propagation of the scalaron is suppressed. This mechanism supports the validity of GR and Newtonian theory ever tested in ground-based experiments, 
whereas the plausible features of the scalaron field, such as driving cosmic expansion, are realized in vacuum regions.

We note that the screening mechanism was often discussed in the presence of non-relativistic matters $T = -\rho$,
where the trace of the energy-momentum tensor is negative $T < 0$.
However, the matter inside neutron stars cannot be described merely by the non-relativistic matters, 
and the corresponding energy-momentum tensor should be written by $T=-\rho +3p$.
When the pressure becomes dominant near the core of the neutron star, the trace of energy-momentum tensor can be positive $T>0$, 
and the possible restoration of the conformal symmetry implies $T \rightarrow 0$ inside the neutron star~\cite{Haensel:2007yy}.
A similar argument can be found in the study of the screening mechanism and the scalaron dynamics in the early Universe~\cite{Erickcek:2013oma, Katsuragawa:2018wbe, Chen:2022zkc}, 
where the scalaron couples with the hot and dense environment in analogy to the neutron star. 
In the current work, we investigate the $T$-dependent potential and its roles inside the neutron star with varying $T$.

The chameleon potential $V_{\mathrm{eff}}(\Phi, T)$ in the $R^2$ gravity~\eqref{eq:R^2} is found to be
\begin{align}
    V_{\mathrm{eff}}(\Phi, T)
    = \frac{1}{12\alpha} \qty(\Phi-\Phi_{\min})^2 - \frac{\Phi_{\min}^2}{12\alpha}\, ,
    \quad \qty(\Phi_{\min} = 1-2\kappa^2 \alpha T)
    \label{eq:cham_pot_R^2_Jordan}
\end{align}
up to a constant term. 
The plots of the potential for several values of $\alpha$ and $T$ are shown in~\figref{fig:chame_pot_JO}.
As the figures show, the potential minimum moves depending on the value of the energy-momentum $T$, and the potential shape becomes steeper as $|\alpha|$ decreases. 
As we will mention later, the above behavior causes difficulties in the numerical calculation for the static system.
The stationary condition $\partial V_{\mathrm{eff}}/\partial \Phi=0$ reduces to 
\begin{align}
    \pdv{V_{\mathrm{eff}}}{\Phi} \qty(\Phi_{\min}, T)
    = \frac{1}{3} \qty[ R(\Phi_{\min}) + \kappa^2 T]
    =0\, .
\end{align}
Thus $R = -\kappa^2 T$ that holds automatically in the GR becomes the condition for the potential minimum. 
In other words, there is no difference between GR and the $R^2$ gravity if the scalaron field always stays at the potential minimum. 
The excitation from the bottom causes nontrivial profiles of the spacetime rather than GR ones. 

The chameleon mass \eqref{eq:cham_mass_Jordan} in the $R^2$ gravity is read as the coefficient of $\Phi^2$, 
\begin{align}
    m_\Phi^2 = \frac{1}{6\alpha} \, .
    \label{eq:cham_mass_R2}
\end{align}
The chameleon mass in the Jordan frame is constant in the $R^2$ model, and the scalaron field merely behaves as a massive particle with the mass $m_\Phi \sim \alpha^{-1/2}$ for positive $\alpha$. 
For negative $\alpha$, the mass $m_{\Phi}$ becomes pure imaginary, and the scalaron field behaves as a tachyon. 
Tachyonic behavior means that the dynamical perturbation grows exponentially, as we can find from the field equation \eqref{eq:EOM_scalar_BD_R^2}. 
Thus the time evolution of the scalaron field causes instability. 

Here we comment on the screening mechanism in the $R^2$ gravity. 
The chameleon mass in the $R^2$ gravity is constant, as is found in Eq.~\eqref{eq:cham_mass_Jordan} in the Jordan frame.
Therefore the screening is absent in this theory. 
In the following analysis, the $T$-dependent behavior of the scalaron field, not the ordinary sense of the screening mechanism, plays an important role.


\section{Static and Spherically Symmetric Star under \texorpdfstring{$R^2$}{R^2} Gravity \label{sec:system}}

\subsection{Setting of the system \label{sec:setting}}

We deal with the static and spherically symmetric star composed of the perfect fluid with a certain EOS in the Jordan frame. 
The geometry inside the star is assumed to be a hydrostatic solution
\begin{align}
    ds^2 = g_{\mu\nu} dx^{\mu} dx^{\nu}
    =- \e^{2\nu (r)} dt^2 + \e^{2\lambda (r)} dr^2 
    + r^2 d\Omega^2\, ,
    \label{eq:metric_assum_in}
\end{align}
where $d\Omega^2 = \bar{g}_{ij} dx^{i} dx^{j} = d\theta^2 + \sin^2\theta \, d\varphi^2$ is the metric on two dimensional sphere $S^2$. 
The matter component is described by the EOS $p=p(\rho)$, and the energy-momentum tensor is defined as
\begin{align}
    T_{\mu\nu} = \qty(\epsilon(r) + p(r)) \, u_{\mu} u_{\nu} 
    + p (r) \, g_{\mu\nu} \, ,
    \label{eq:EMtensor}
\end{align}
with rest mass density $\rho(r)$, total energy density $\epsilon(r)$~\footnote{Note that this system is highly relativistic so that rest mass density $\rho(r)$ and total energy density $\epsilon(r)$ should be rigorously distinct. 
In this work, we often use $\rho$ as the parameter of the system, while the actual calculation is mostly based on $\epsilon$.}
, pressure $p(r)$, and 4-velocity of static fluid $u^{\mu}$. 
It should be noted that we work on the Jordan frame, and the above quantities do not contain the contribution from the scalaron field but only contain that from the matter. 
The center of the star is placed at $r=0$, while the surface radius $r_s$ of the star is defined as the radius where the fluid pressure vanishes $p(r_s) = 0$.

We assume the external region of the star $r>r_s$ is a vacuum, and the cosmological constant vanishes. 
In the GR case, this assumption tells that the outer geometry must be the Schwarzschild solution due to Birkhoff-Jebsen's theorem \cite{birkhoff1923relativity, jebsen2005general}. 
However, this theorem is generally absent in the $F(R)$ gravity, and the outer solution is not necessarily the Schwarzschild one, as mentioned in the previous section. 
Furthermore, it was found in \cite{Numajiri:2021nsc} that the possible lowest-order modification in the $F(R)$ should be the noninteger power term of curvature, 
\begin{align}
    F(R) = R + a R^{b}\, , 
    \quad \qty(a\in \mathbb{R}, \; 1 < b < 2)
\end{align}
to realize the compact star system whose outer spacetime is the Schwarzschild one (under the assumption of the polytropic EOS).
The contraposition of this statement suggests that the external geometry in the $R^2$ gravity is expected to have a nontrivial curvature distribution.
Because the curvature is related to the scalaron field, the $R^2$ gravity may show the nonvanishing scalaron field surrounding the compact star, which could be called the scalarized solution. 

We consider the metric for outer spacetime to have the same form as that for the internal one~\footnote{We use the same function as the internal region because they should be connected with junction conditions we will see later.};
\begin{align}
    ds^2 = g_{\mu\nu} dx^{\mu} dx^{\nu}
    =- \e^{2\nu (r)} dt^2 + \e^{2\lambda (r)} dr^2 
    + r^2 d\Omega^2\, .
    \label{eq:metric_assum_out}
\end{align}
Hereafter, we analyze the asymptotic behaviors of the metric components from those of the scalaron or the curvature. 
According to the discussion on the chameleon potential in the previous section, the scalaron field $\Phi$ behaves as a free massive particle in the $R^2$ gravity:
\begin{align}
    \Box \Phi = m_{\Phi}^2 \qty(\Phi-\Phi_{\min})\, .
    \quad \qty(m_{\Phi}^2 = \frac{1}{6\alpha}, \;
    \Phi_{\min} = 1-2\kappa^2 \alpha T)
    \label{eq:scalarEOM_eff}
\end{align}
Assuming the asymptotically flat spacetime at the external region and $T=0$, the asymptotic solution of Eq.~\eqref{eq:scalarEOM_eff} would be 
\begin{align}
    \Phi (r) -1 \sim \frac{c_1}{r} \e^{m_{\Phi} r} 
    + \frac{c_2}{r} \e^{- m_{\Phi} r} \, .
    \label{eq:asym_sol}
\end{align}
Eq.~\eqref{eq:asym_sol} shows the decaying and growing modes in space, reflecting the static spacetime in Eq.~\eqref{eq:metric_assum_out}.
For the assumption of the asymptotic flat spacetime, it is relevant to consider the decaying mode, and thus $\Phi \rightarrow 1$.
Moreover, Eq.~\eqref{eq:scalar_BD_R2} suggests $\Phi \rightarrow 1$ corresponds to the limit to GR and $R\rightarrow0$.
Therefore, the deviation from the GR itself is expected to be the exponentially decreasing:
\begin{align}
    \Phi (r) -1 \propto \frac{1}{r} \e^{- m_{\Phi} r} \, .
    \label{eq:phi_boundary}
\end{align}
In other words, compact stars in the $R^2$ gravity have exponentially decaying scalar hair. 
From the equivalence of the curvature and the scalaron field, the curvature 
is also predicted to decay exponentially with some typical length, which is of the same order as the Compton length of the scalaron field $m_{\Phi}^{-1}=\sqrt{6\alpha}$. 

Because the deviation from GR is decreasing in asymptotic flat spacetime, we can conclude that Birkhoff-Jebsen's theorem in GR is restored further away from the Compton wavelength $r \gg \sqrt{\alpha}$ and that the spacetime solution is asymptotically Schwarzschild one;
\begin{align}
    \lim_{r\rightarrow \infty} \e^{2\nu(r)} 
    = \lim_{r\rightarrow \infty} \e^{-2\lambda(r)}
    = \qty(1-\frac{2M}{r})\, ,
    \quad 
    \lim_{r\rightarrow \infty} R(r) 
    = \lim_{r\rightarrow \infty} R'(r) = 0\, ,
    \label{eq:asym_Sch_cond}
\end{align} 
($R(r), R' (r)$: the curvature and its first derivative with respect to the radial coordinate $r$) because of its observational viability. 
The constant $M$ corresponds to the Schwarzschild mass (or ADM mass) for an infinite-distant observer; hence, $M$ is the mass we observe. 
Later we will show the mass-radius relation of the stars using this definition. 
The metric component function $\nu(r), \lambda(r)$ and the curvature $R(r)$ (or the scalaron field distribution $\Phi(r)$) are to be calculated numerically with these asymptotic conditions and the continuity conditions at the surface we will mention later.

Here we comment on the discussions in some existing works. 
In Ref.~\cite{Ganguly:2013taa}, the authors have claimed that the neutron stars have no hair in the $R^2$ gravity based on the black hole no-hair theorem proved in this gravitational theory~\cite{Whitt:1984pd, Mignemi:1991wa}. 
However, their proofs of the no-hair theorem strongly depend on the black hole features such as vacuum and the existence of the horizon. 
For the compact stars, they consist of vacuum and non-vacuum regions with properly connected geometries.
Although the existing works have also considered the non-vacuum case, their analyses can apply only to the case where the trace of the energy-momentum tensor of matter is vanishing (e.g. the electromagnetic field).
In a realistic consideration for the compact stars, the trace of energy-momentum tensor is non-vanishing,
and thus it is still to be analyzed deliberately whether the no-hair theorem also holds in such a case.
\footnote{
The no-hair theorem is proved in some particular cases: Compact star system with reflecting surface is discussed in \cite{Hod:2016vkt, Bhattacharjee:2017huz, Peng:2019iac}. And also, the case with shift-symmetric Horndeski theory is proved in \cite{Lehebel:2017fag}. 
Their assumptions do not coincide with our consideration, and thus no contradiction appears.}

Moreover, the results in some previous works \cite{Yazadjiev:2014cza, Astashenok:2018iav, Nava-Callejas:2022pip} and ours show the existence of haired solution numerically. 
They imply it is suspicious if such a no-hair theorem holds even in compact stars.

In Refs.~\cite{Resco:2016upv, Feola:2019zqg}, they chose $\alpha<0$ to realize the radial perturbative stability of the system. 
However, this choice of the parameter $\alpha$ leads to the tachyonic scalar mode, as is apparent in Eq.~\eqref{eq:cham_mass_R2}. 
Therefore the scalaron field equation \eqref{eq:scalarEOM_eff} suggests that the dynamical (time-dependent) perturbation behaves exponentially and causes instability. 
Also, even under the static configuration, the stellar mass of the system tends to be infinite because of the spatial oscillation in the geometry~\cite{Astashenok:2018iav}. 
This case of the negative $\alpha$ is unphysical from the observational point of view. 
For these reasons, the case of negative $\alpha$ is not plausible in the actual compact star configurations.

We also mention the junction conditions in this system. 
For all over the system, we assume that there is no delta-function-like discontinuity for the matter field,
\begin{align}
    \qty[T_{\mu\nu}]=0\, .
    \label{eq:junc_T}
\end{align}
Here $\qty[\cdots]$ denotes the discrepancy over some timelike hypersurface. 
In the $F(R)$ gravity theory, which contains additional scalar DOF, the junction conditions are not only the Israel condition~\footnote{The continuity condition for the extrinsic curvature $\qty[K_{\mu\nu}]=0$ is automatically guaranteed by the continuity of $h_{\mu\nu}$ and $T_{\mu\nu}$.}
\begin{align}
    \qty[h_{\mu\nu}]=0\, ,
    \label{eq:junc_h}
\end{align}
but also the curvature continuity conditions \cite{Deruelle:2007pt}
\begin{align}
    \qty[R] = 0\,, \quad \qty[\nabla_{\mu} R]=0\, ,
    \label{eq:junc_R}
\end{align}
where $h_{\mu\nu}$ is the induced metric for the hypersurface. 
We demand these conditions when we connect the numerical inner- and outer solutions, which are independently calculated.

By taking into account the junction condition at the surface $r=r_s$, this system can be recognized as a two-boundaries-value problem for the center $r=0$ and the spacelike infinity $r\rightarrow \infty$. 
For such a system, we usually choose the shooting method from the center as the first choice to solve them. 
However, as mentioned above, this system has exponentially decaying scalar hair, and it causes difficulty in finding a consistent solution numerically.
To obtain the asymptotically flat solution, we would like to pick up the second term of the general solution \eqref{eq:asym_sol}. 
However, the first term blows up and prevents convergence during numerical iterative integration. 
Such a divergence becomes fatal for integrating longer distances and making $\alpha$ smaller, and it is almost impossible to solve the system with the usual one-way shooting method starting from the center. 

Therefore we choose to shoot from both boundaries, i.e., from the center and the (artificial) endpoint $r_e$ with a sufficiently large distance. 
Demanding that the obtained two numerical solutions satisfy the junction conditions we just mentioned, we find consistent asymptotic flat solutions. 
This method is also essential to solve this system. We will explain the details of this setting in \secref{sec:implementation}.

\subsection{Modified TOV Equations}
We now derive the modified version of the celebrated Tolman-Oppenheimer-Volkoff (TOV) equation in the $R^2$ gravity based on the settings we prepared in the last part. 
The geometries inside and outside the star are assumed to be static and spherically symmetric solutions \eqref{eq:metric_assum_in}, \eqref{eq:metric_assum_out}. 
The nonvanishing geometrical quantities (the Levi-Civita connection, the Riemann tensor, the Ricci tensor, and the Ricci scalar) for this spacetime are as below,
\begin{align}
    &\Gamma^r_{tt} = \e^{-2(\lambda - \nu)}\nu' \, ,
    \quad \Gamma^t_{tr}=\Gamma^t_{rt}=\nu'\, , \quad 
    \Gamma^r_{rr}=\lambda'\, , \nn
    &\Gamma^i_{jk} = \bar{\Gamma} ^i_{jk}\, ,\quad 
    \Gamma^r_{ij}=-\e^{-2\lambda}r \bar{g}_{ij} \, ,
    \quad \Gamma^i_{rj}=\Gamma^i_{jr}=\frac{1}{r} \, \delta^i_{j} \, , \\
    &R_{rtrt} = \e^{2\nu}\left[ \nu'' + \left(\nu' - \lambda'\right)\nu' \right] \, ,\quad 
    R_{titj} = r\nu'\e^{2(\nu - \lambda)} \bar{g}_{ij} \, , \nn
    &R_{rirj} = \, \lambda' r \bar{ g}_{ij} \, , \quad
    R_{ijkl} = \left( 1 - \e^{-2\lambda}\right) r^2
    \left(\bar{g}_{ik} \bar{g}_{jl} - \bar{g}_{il} \bar{g}_{jk} \right)\, ,\\
    &R_{tt}= \e^{2\left(\nu - \lambda\right)} \left[
    \nu'' + \left(\nu' - \lambda'\right)\nu' + \frac{2\nu'}{r}\right] \, ,\quad 
    R_{rr} = - \left[ \nu'' + \left(\nu' - \lambda'\right)\nu' \right] 
    + \frac{2 \lambda'}{r} \, , \nn
    &R_{ij} = \left[ 1 + \left\{ - 1 - r \left(\nu' - \lambda' \right)\right\}\e^{-2\lambda}\right]\bar{g}_{ij}, \\
    &R= \e^{-2\lambda}\left[ - 2\nu'' - 2\left(\nu'
    - \lambda'\right)\nu' - \frac{4\left(\nu'- \lambda'\right)}{r} + \frac{2\e^{2\lambda} - 2}{r^2} \right] \, .
    \label{eq:cuvatures}
\end{align}
The matter inside the star is assumed to be the perfect fluid with the specified EOS $p=p(\rho)$, whose energy-momentum tensor is given in Eq.~\eqref{eq:EMtensor}. 
This energy-momentum tensor satisfies the conservation law due to the generalized Bianchi identity of the $F(R)$ gravity in the Jordan frame.
\begin{align}
    \nabla^\mu T_{\mu \nu} = 0\, .
    \label{eq:continuity}
\end{align}

Then the nontrivial components of the field equation for the metric \eqref{eq:EOM_Jordan} are found to be
\begin{align}
    \label{eq:EOM00}
    &- \frac{1}{2} F - \e^{- 2 \lambda} \left\{
    \nu'' + \left(\nu' - \lambda'\right)\nu' + \frac{2\nu'}{r}\right\} F_R 
    + \e^{ -2\lambda} \left( F_R'' + \left( - \lambda' + \frac{2}{r} \right) F_R' \right) 
    =\, - \kappa^2 \epsilon \, ,\nn
    \\
    \label{eq:EOM11}
    &\frac{1}{2} F + \e^{ -2\lambda} \left\{ \nu'' + \left(\nu' - \lambda'\right)\nu' 
    - \frac{2 \lambda'}{r} \right\} F_R - \e^{ -2\lambda} 
    \left( \nu' + \frac{2}{r} \right) F_R' =\, - \kappa^2 p \, ,\\
    \label{eq:EOM22}
    &\frac{1}{2} F - \frac{1}{r^2} \left\{ 1 + \left\{ - 1 - r \left(\nu' 
    - \lambda' \right)\right\}\e^{-2\lambda}\right\} F_R 
    - \e^{-2\lambda} \left( F_R'' + \left( \nu' - \lambda' + \frac{1}{r} \right) F_R' \right) 
    =\, - \kappa^2 p\, .
\end{align}
Using the expression of the curvature in \eqref{eq:cuvatures}, we can rewrite the above equations \eqref{eq:EOM00}--\eqref{eq:EOM22} as
\begin{align}
     \lambda'
    =&\, 
    \frac{ \e^{2 \lambda} \left\{ r^2 
    \left( 2 \kappa^2 \epsilon - F(R) \right) 
    +  F_R \left(r^2 R - 2\right)\right\}
    + 2 r^2 F^{(3)}_{R} \left(R'\right)^2 
    + 2 r F_{RR} \left( r R'' + 2 R' \right) + 2 F_R }
    {2 r \left(2 F_R + r F_{RR} R'\right)} \, ,
    \label{eq:lambda'_gene}\\
    \nu'
    =&\, 
    \frac{\e^{2\lambda} 
    \left\{r^2 \left(2 \kappa^2 p + F(R)\right) - F_R \left( r^2 R - 2 \right) \right\} 
    - 2 \left( 2 r F_{RR} R' + F_R \right)}
    {2r \left( 2F_R + rF_{RR} R'\right)},
    \label{eq:nu'_gene}\\
    R^{\prime \prime}
    =&\, \frac{F_R}{F_{RR}}
    \left[\frac{1}{r}\left(3 \nu^{\prime}-\lambda^{\prime}+\frac{2}{r}\right)
    +\mathrm{e}^{2 \lambda} \left(\frac{1}{2} R-\frac{2}{r^{2}}\right)\right]
    +\left(\lambda^{\prime}+\frac{1}{r}\right) R^{\prime}
    -\frac{F^{(3)}_{R}}{F_{R R}} R^{\prime 2} \, .
    \label{eq:R''_gene}
\end{align}
Also the conservation law \eqref{eq:continuity} reduces to 
\begin{align}
    0 = (\epsilon + p) \nu' + p'\, .
    \label{eq:continuity_TOV}
\end{align}
The Eqs.~\eqref{eq:lambda'_gene}--\eqref{eq:continuity_TOV} are the modified TOV equations for an internal region in general $F(R)$ gravity. 
Those for the external region are given just by putting matter quantities $\epsilon, p$ to vanish. 
Here we use curvature $R(r)$ as an independent field, which corresponds to scalar DOF in the scalar-tensor description. 
Note that using Eq.~\eqref{eq:cuvatures}, we can delete the field equation for $R$ \eqref{eq:R''_gene}, but the field equation for the metric components becomes higher-order differential equations.

In the $R^2$ gravity Eqs.~\eqref{eq:lambda'_gene}--\eqref{eq:R''_gene} become
\begin{align}
    \lambda' &= \frac{ \e^{2 \lambda} 
    \left\{ 2 \kappa^2 r^2 \epsilon + \alpha R (r^2 R 
    -4 ) -2 \right\}
    + 4 \alpha   \left( r^2 R'' + 2 r R' + R \right) + 2}
    {4 r \left( \alpha (r  R' + 2 R) + 1\right)} \, , 
    \label{eq:lambda'_R2}\\
    \nu' &= \frac{ \e^{2 \lambda} 
    \left\{ 2 \kappa^2 r^2 p - \alpha R (r^2 R 
    -4 ) + 2 \right\}
    - 2  \left( 4 \alpha r R' + 2 \alpha R + 1 \right) }
    {4 r \left( \alpha (r  R' + 2 R) + 1\right)} \, , 
    \label{eq:nu'_R2}\\
    R^{\prime \prime}
    & =\frac{1 + 2 \alpha R}{2 \alpha }
    \left[\frac{1}{r}\left(3 \nu^{\prime}-\lambda^{\prime}+\frac{2}{r}\right)
    +\mathrm{e}^{2 \lambda} \left(\frac{1}{2} R-\frac{2}{r^{2}}\right)\right]
    +\left(\lambda^{\prime}+\frac{1}{r}\right) R^{\prime} \,.
    \label{eq:R''_R2}
\end{align}
These and Eq.~\eqref{eq:continuity_TOV} are the equations to be solved numerically if the equation of states (EOS) $p=p(\epsilon)$ is given. 
When numerical integration is performed, we make them dimensionless properly using
\begin{gather*}
    M_g = M_\odot \simeq 1.99 \cdot 10^{33} \: \si{g}, \quad 
    r_g = \frac{G M_\odot}{c^2} \simeq  1.48 \cdot 10^{5} \: \si{cm},\\
    \rho_g = \frac{M_g}{r_g^3} \simeq 6.18 \cdot 10^{17} \: \si{g\,cm^{-3}},\quad
    p_g = \frac{M_g c^2}{r_g^3} \simeq 5.55 \cdot 10^{38} \: \si{g\,cm^{-1}\,s^{-2}}.
\end{gather*}

\subsection{Boundary Conditions and Asymptotic Solutions}

Here we need to find the concrete asymptotic forms of solutions from the modified TOV equations. 
Around the center of the star $r=0$, we require 
\begin{align}
    \lambda (0) = R'(0) = 0\, ,
    \label{eq:BC_0}
\end{align}
to avoid the conical singularity. 
The other quantities take some constant values, 
\begin{align}
    \nu(0)=\nu_0, \quad
    R(0)=R_0, \quad
    p(0)=p_0.
\end{align}
Here $p_0$ is given by hand as a parameter. 
The constants $\nu_0$ and $R_0$ are determined by requiring that the solutions should satisfy the asymptotically flat conditions at $r \rightarrow \infty$. 
The equations \eqref{eq:lambda'_gene}--\eqref{eq:R''_gene} are not available at $r = 0$, and hence it is better to construct asymptotic solutions around the center. 
By expanding Eqs.~\eqref{eq:lambda'_gene}--\eqref{eq:R''_gene} around $r=0$, the solutions are found to behave as
\begin{gather}
    \lambda(r) \sim \frac{1}{2} \lambda_2 r^2,  \quad
    \nu(r) \sim  \nu_0 + \frac{1}{2} \nu_2 r^2, \nn
    R(r) \sim R_0 + \frac{1}{2} R_2 r^2, \quad
    p(r) \sim p_0 + \frac{1}{2} p_2 r^2,
    \label{eq:asym_center}
\end{gather}
with
\begin{align*}
    &\lambda_2 = \frac{1}{18(2\alpha R_0 + 1)}
    \qty[2\kappa^2 (2\epsilon_0 + 3 p_0 ) + \qty(3\alpha R_0 +2)R_0],\nn
    &\nu_2 = \frac{1}{18(2\alpha R_0 + 1)}
    \qty[2\kappa^2 (2\epsilon_0 + 3 p_0 ) - \qty(3\alpha R_0 +1)R_0] ,\nn
    &R_2 = \frac{1}{18\alpha} 
    \qty[\kappa^2 (-\epsilon_0 + 3p_0) +  R_0], \nn
    &p_2 
    = - \qty(\epsilon_0 + p_0)\nu_2  ,
\end{align*}
where $\epsilon_0$ is obtained by solving the EOS for given $p_0$.

At the infinitely distant boundary $r \rightarrow \infty$, we demand the geometry is asymptotically Schwarzschild solution in Eq.~\eqref{eq:asym_Sch_cond}. 
The Schwarzschild mass $M$ for $r\rightarrow \infty$ is the stellar mass observed by an infinitely distant observer. 
Thus, this mass should be recognized as the observable that appears in the mass-radius relation and other measurements. 
On the other hand, we can define the effective Schwarzschild mass $m(r)$ inside the 2-sphere with radius $r$:
\begin{align}
    m(r) \equiv \frac{r}{2} \qty(1 - \e^{-2\lambda (r)})\, .
\end{align}
In GR configuration, this mass corresponds to the stellar mass at the surface; that is, $m(r_s)=M$. 
The situation is, however, different in the $R^2$ gravity. 
Due to the nontrivial geometry outside the star (i.e., the existence of the scalar hair), $m(r)$ is not equal to $M$ for finite $r$ but for the limit $r \rightarrow \infty$. 
In \secref{sec:sub_results}, this fact will be shown explicitly with numerical results.

The asymptotic flat solutions for large $r$ can be obtained by solving Eqs.~\eqref{eq:lambda'_gene}--\eqref{eq:R''_gene} for $r \gg M$ and imposing consistency with Eq.~\eqref{eq:cuvatures}:
\begin{align}
    &\lambda (r) \sim -\frac{1}{2} \log \qty(1- \frac{2M}{r})
    - 6 \alpha C \, r \, K_1 \qty(\frac{r}{2\sqrt{\alpha}})\, , \nn
    &\nu (r) \sim \frac{1}{2} \log \qty(1- \frac{2M}{r})
    - 2 \alpha C \, r \, K_1 \qty(\frac{r}{2\sqrt{\alpha}})\, , \nn
    &R(r)  \sim C \, r \, K_1 \qty(\frac{r}{2\sqrt{\alpha}})\, .\quad
    \label{eq:asym_end}
\end{align}
Here $C$ is a constant to be determined to satisfy the asymptotic conditions in a consistent manner. 
$K_n(x)$ is the modified Bessel function of the second kind. 
As we can see, the deviations from the Schwarzschild solution have the form of exponentially decaying terms depending on the typical length, which is proportional to the Compton length of the scalaron $m_{\Phi}^{-1} = 1/\sqrt{6\alpha}$. 
This result is consistent with the discussion on the scalaron field profile in Eq.~\eqref{eq:phi_boundary} and indicates the existence of the contribution from the scalar hair outside the star.

Now let us check the consistency of the number of equations and the conditions. 
The equations we use consist of four differential equations; three first-order ordinary differential equations (ODEs) \eqref{eq:lambda'_R2}, \eqref{eq:nu'_R2}, and \eqref{eq:continuity} for the metric components $\lambda, \nu$ and the pressure $p$, and one second-order ODE \eqref{eq:R''_R2} for the curvature $R$ as the additional scalar DOF.
Therefore we need to specify five boundary conditions, which are given by
\begin{align}
    \lambda(0)=0, \quad
    \nu(r) \xrightarrow{r\rightarrow \infty} 0, \quad
    R(r) \xrightarrow{r\rightarrow \infty} 0, \quad
    R'(0) = 0, \quad
    p(0) = p_c\, .
\end{align}
Here $p_c$ (or $\rho_c, \epsilon_c$) serves as the free parameter to be fixed by hand. 
The asymptotic solutions in  Eq.~\eqref{eq:asym_center} around the center $r=0$ are characterized by two parameters $\nu_c = \nu(0)$ and $R_c=R(0)$. 
In the sense of the usual shooting method, all we have to do is find the appropriate $\nu_c, R_c$, which satisfies all of the boundary conditions. 
Due to the numerical difficulty that we mentioned in \secref{sec:setting}, however, we choose to solve the system from both of the boundaries simultaneously and try to find the parameters $M$ and $C$ which characterize the asymptotic solutions \eqref{eq:asym_end} for $r\rightarrow \infty$ in addition.


\section{Numerical Results and Discussions\label{sec:results}}

\subsection{Implementation \label{sec:implementation}}

\begin{figure}
    \centering
    \includegraphics[keepaspectratio, width=\linewidth]{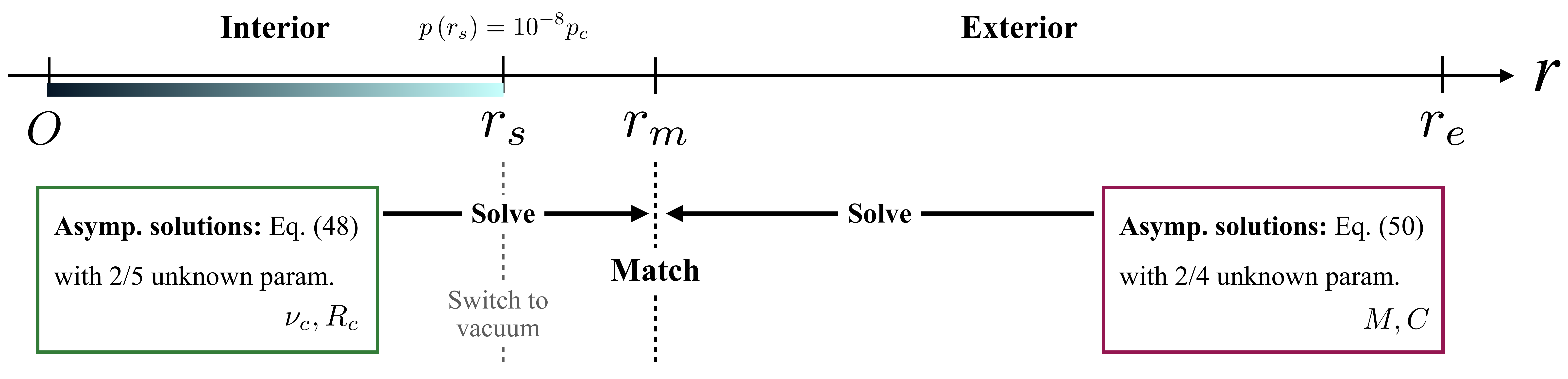}
    \caption{Illustration of the scheme for numerical integration. The multi-boundaries shooting method is used to deal with the diverging numerical instabilities.}
    \label{fig:implement}
\end{figure}

We now explain the numerical setting we use (see also \figref{fig:implement}). As we mentioned in \secref{sec:setting}, the system with Eqs.~\eqref{eq:continuity_TOV}--\eqref{eq:R''_R2} should be solved as the multi--boundaries values problem where boundaries are placed at $r=0$ and $r\rightarrow \infty$. Due to the exponentially behaving scalar DOF, this system is numerically {\it unstable} so that the usual one-way shooting method is unavailable.

For the above reasons, we use the shooting method from both boundaries. 
Details are as below: Firstly, the two boundaries were set as $r_c = 10^{-3} r_g$ for the center and $r_e = 10^2 \cdot \sqrt{\alpha}$ for the sufficiently distant boundary. 
The factor $\sqrt{\alpha}$ in $r_e$ is necessary to guarantee that the distance of the endpoint from the center is much larger than the Compton length of the scalaron field. 
Moreover, the surface of the star $r=r_s$ was numerically defined as a surface where $p(r_s) \leq 10^{-8} \, p_c$ is satisfied.
The matching point $r_m$ was placed at $r_m = 10 \, r_g$, which is slightly further than the typical surface radius of the stars~\footnote{
Here we would like to comment on the values of $r_i$ and $r_e$. The value of $r_i$ cannot be lengthened but shortened, because of numerical instabilities. In particular, it should be made shorter flexibly for small $\alpha(<1)$, in which the numerical instabilities are more likely to appear. In that case, we must care about whether $r_s < r_i$ still holds. For the value of $r_e$, we can extend by $\sim 1.5 \cdot 10^2\sqrt{\alpha}$ so that $M$ and $r_s$ vary $\sim\order{10^{-4}}$. The numerical convergence becomes worse if we take a much longer value for $r_e$, because the value of $C$ becomes much larger than other parameters.}. 

Under the above configuration, we solve Eqs.~\eqref{eq:continuity_TOV}--\eqref{eq:R''_R2} numerically from both boundaries $r_c, \,r_e$ toward the matching point $r_m$. 
For the center $r_c$, the asymptotic solutions \eqref{eq:asym_center} are firstly assumed with two unspecified boundary values $\nu_c, \, R_c$ and a given parameter $p_c$. 
We solve the equations numerically with a given EOS until they reach the surface $r = r_s$ and switch them to the vacuum ones where $\epsilon,\, p$, and their derivatives vanish. 
At the surface of the star, $\lambda,\nu, R,$ and $R'$ have to be continuous because of the junction conditions \eqref{eq:junc_T}--\eqref{eq:junc_R}. 
The integration is performed until the matching point $r=r_m$. 

On the other hand, at the endpoint $r=r_e$, the asymptotic solution Eq.~\eqref{eq:asym_end} is assumed with two unspecified constants $M$ and $C$. 
The vacuum version of the modified TOV equations \eqref{eq:lambda'_R2}--\eqref{eq:R''_R2} are solved "backward" from $r=r_e$ to $r=r_m$. 
At the matching point $r_m$, we impose the continuity condition on $\lambda, \nu, R,$ and $R'$ that difference between the two calculated solutions is order of $10^{-4}$ of solution ($\Delta\lambda(r_m) < 10^{-4} \lambda(r_m)$ for example). 
This condition leads to determine constants $\nu_c, R_c, M,$ and $C$ which specify the asymptotic solution \eqref{eq:asym_center} and \eqref{eq:asym_end}. 
By performing this procedure for each value of the central pressure $p_c$, we obtain the family of solutions and the mass-radius relation corresponding to the EOS we assumed. 

Here we comment on the EOS. As EOS $p=p(\rho)$, we used the piecewise form \cite{Read2008iy} of SLy EOS \cite{Alford:2004pf}, which corresponds to the neutron star with a quark matter core with QCD correction, and APR4 EOS \cite{Douchin:2001sv} which describes the crust and the liquid core of nuclear matters based on the $N$-body simulation. 
It should be noted that some conditions on the upper value for $p_c$ exist:
\begin{inparaenum}[(i)]
    \item Condition from the EOS itself that states the (square of) sound speed $v^2=\dd{p}/\dd{\epsilon}$ does not exceed light speed $c^2$.
    \item Condition coming from the positivity of the chameleon mass in the Einstein frame in Eq.~\eqref{eq:cond_chamemass}. 
\end{inparaenum}
The infimum of these two bounds should be taken as the upper limit of the region of central pressure $p_c$. 

For the value of the modification parameter $\alpha$, we take $\alpha = 0$, $0.5 r_g^2$, $r_g^2$, $2r_g^2$, $10 r_g^2$, and $100 r_g^2$. These values are compatible with observational constraints such as the Gravity Probe B and the pulsar B in the PSR J0737-3039 system \cite{Naf:2010zy}, while recent investigation on the dynamical stability of compact stars suggests incompatible one \cite{Pretel:2020rqx}. However, we do not compare our results with such constraints in this paper. This is because in principle the parameter $\alpha$ cannot be constrained from only the mass-radius relation in order to another dependence on the choice of EOS \cite{Numajiri:2021nsc}. Thus we concentrate here on constructing proper compact star systems with correct exterior geometry and revealing the qualitative nature of the internal scalaron field.

To perform the numerical integration of the modified TOV equations, we use the BDF method with the adaptive step--size control distributed in Wolfram Mathematica \cite{Mathematica}. 
Some tuning on the accuracy and precision is essential for the calculation when $\alpha$ is small ($\alpha<1$). 
Moreover, there is serious numerical difficulty in the calculation of $\alpha < 0.5 r_g^2$, although the results with such $\alpha$ would be indistinguishable from GR solutions.

\subsection{Results and Discussions\label{sec:sub_results}}

\begin{figure}[tbp]%
  \begin{minipage}[t]{0.5\linewidth}%
    \centering%
    \includegraphics[keepaspectratio, width=0.9\linewidth]{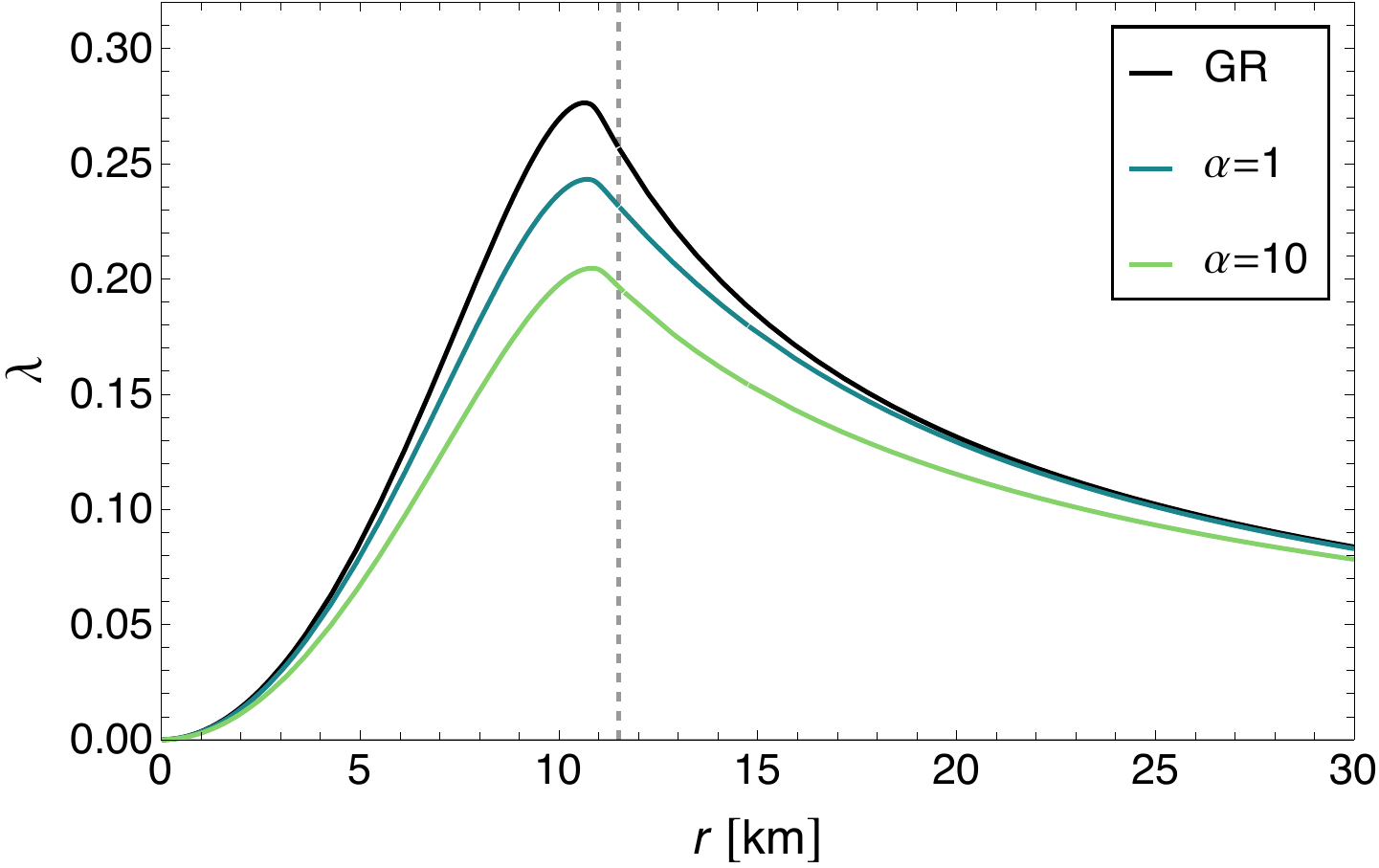}%
    \subcaption{Metric component $\lambda(r)$}%
    \label{fig:L}%
  \end{minipage}%
  \begin{minipage}[t]{0.5\linewidth}%
    \centering%
    \includegraphics[keepaspectratio, width=0.9\linewidth]{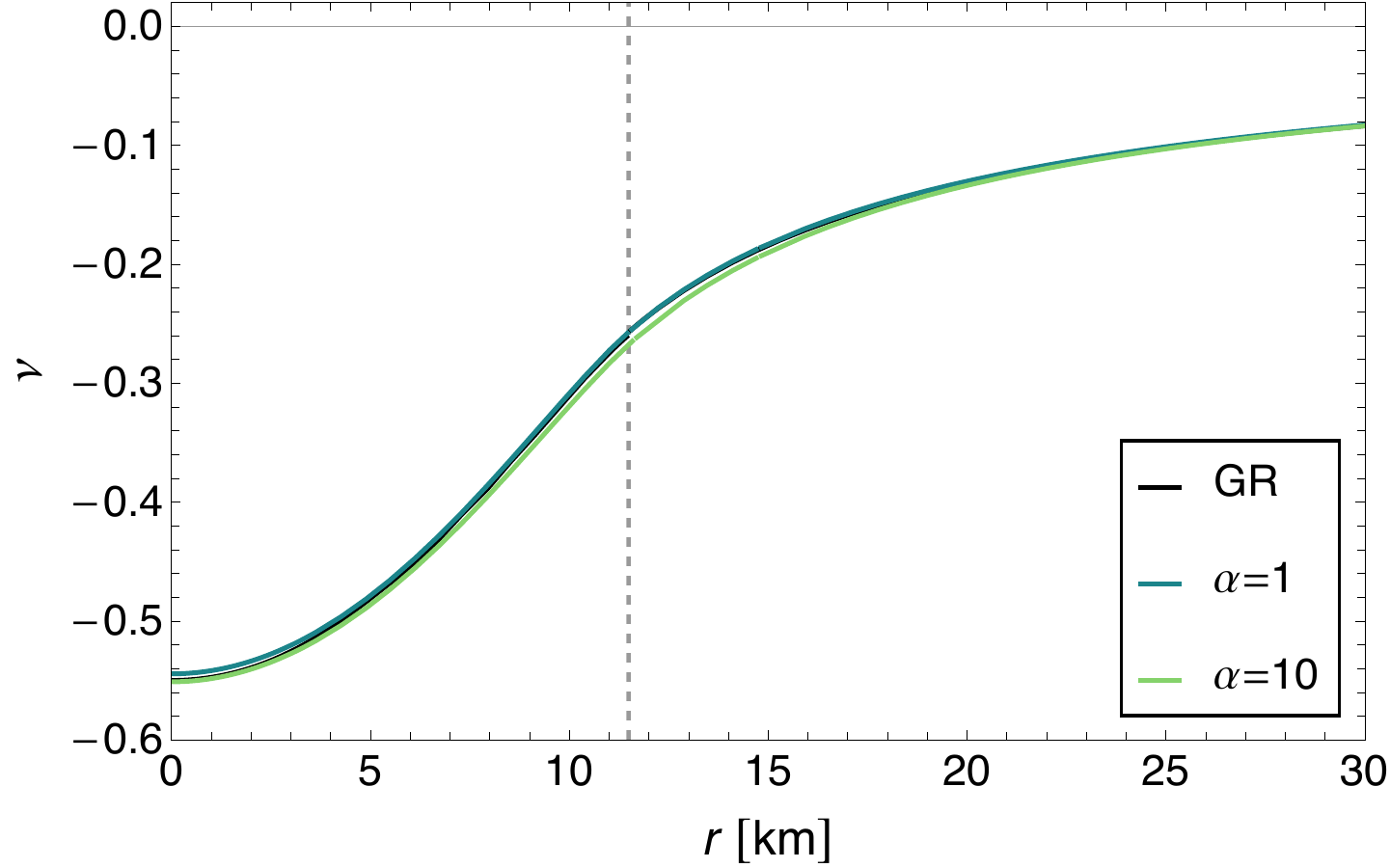}%
    \subcaption{Metric component $\nu(r)$}%
    \label{fig:Nu}%
  \end{minipage}
  \\[5pt]
  \begin{minipage}[t]{0.5\linewidth}%
    \centering%
    \includegraphics[keepaspectratio, width=0.9\linewidth]{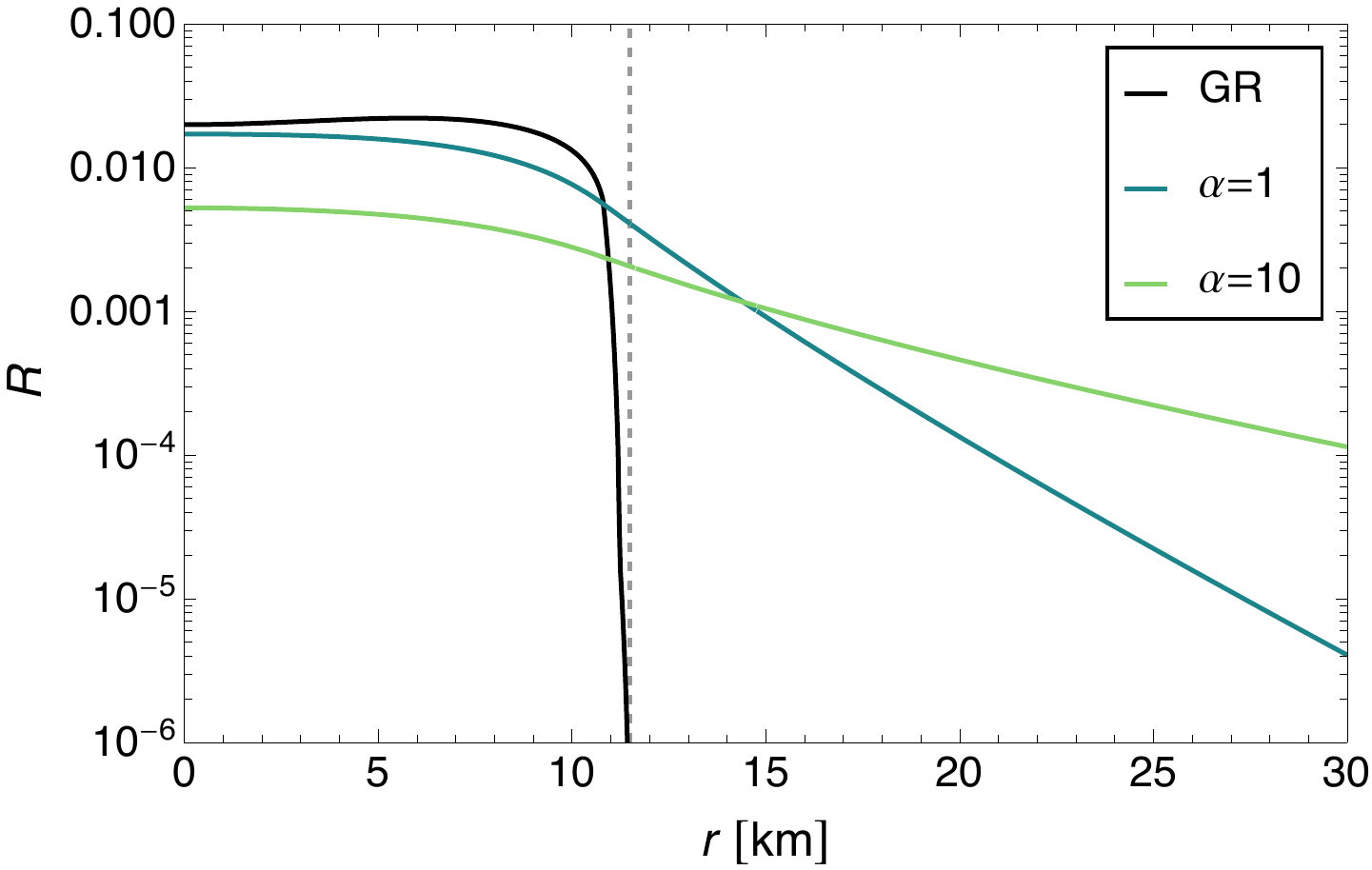}%
    \subcaption{Curvature profile $R(r)$}%
    \label{fig:R}%
  \end{minipage}%
  \begin{minipage}[t]{0.5\linewidth}%
    \centering%
    \includegraphics[keepaspectratio, width=0.9\linewidth]{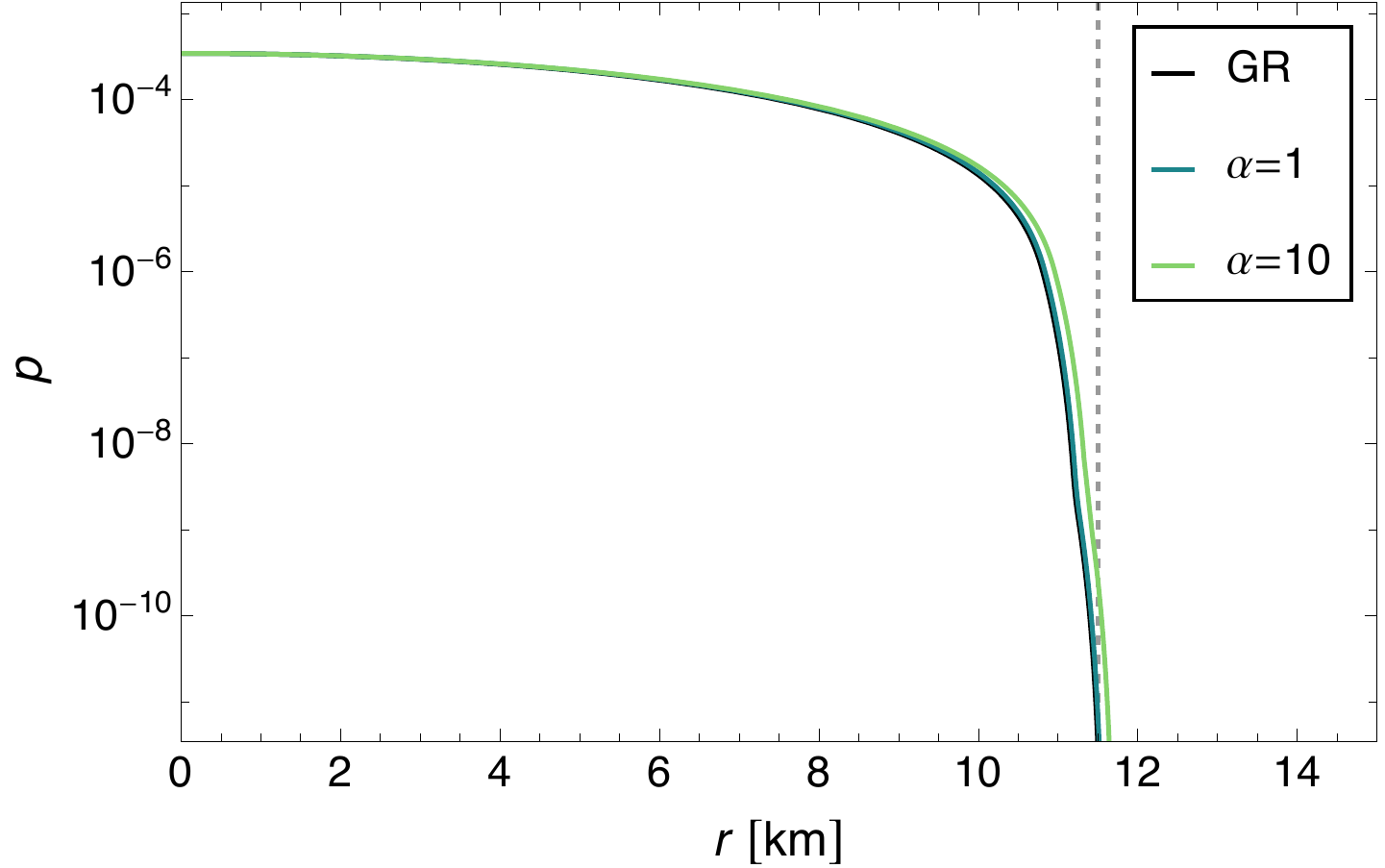}%
    \subcaption{Pressure profile $p(r)$}%
    \label{fig:p}%
  \end{minipage}
  \caption{
The solutions with $\rho_c = 1.00 \cdot 10^{15}\; \si{g/cm^3}$ (SLy EOS) for GR and $\alpha = r_g^2,\,  10\, r_g^2$. 
The vertical dashed line represents the radius $r_s = 11.6 \, \si{km}$ in the GR solution (as a reference value for non--zero $\alpha$ solutions). 
The behaviors of $\lambda(r)$ and $R(r)$ show the large deviations from the GR solution (or Schwarzschild solution) for large $\alpha$, while $\nu(r)$ and $ p(r)$ show small deviations.
}
  \label{fig:geo_SLy}
\end{figure}%

\subsubsection{Geometry}

Firstly we depict the plots of the geometrical quantities $\lambda(r), \nu(r), R(r)$ and the pressure profile $p(r)$ for $\alpha = 0\, (\text{GR}),\,  r_g^2,\, 10\, r_g^2$ in \figref{fig:geo_SLy}. 
The central pressure were taken as all the same value $p_c \simeq 3.44\cdot10^{-4}\, p_g$ (corresponding to the rest mass density $\rho = 1.00 \cdot 10^{15}\; \si{g/cm^3}$) with the SLy EOS. 
The differences in the mass and radius for different $\alpha$ were $\order{10^{-2}}$ so that they are hardly seen in these plots, as we find from the plot for the pressure in \figref{fig:p}. 

In these plots, the metric component $\lambda(r)$ and the curvature $R(r)$ show drastic deviations from GR while $\nu(r)$ and $p(r)$ do not. 
In \figref{fig:R}, the curvature $R(r)$ does not vanish in the external region for nonzero $\alpha$, whereas it vanishes for the $\alpha=0$ (GR) case. 
The decreasing of the curvature becomes softer for large $\alpha$. 
This result indicates the aforementioned scalarization effect in the $F(R)$ gravity.

This property also can be seen from \figref{fig:L}. 
In this plot, all the profiles of the Schwarzschild solutions with corresponding masses $M$ for several values of $\alpha$ are almost the same as the external profile for $\alpha=0$ (GR) due to the mentioned small discrepancies on $M$. 
The deviation from the Schwarzschild solution becomes significant around the surface $r=r_s$ and decreases when $r\rightarrow \infty$, and this deviation becomes larger as we choose larger $\alpha$. 
This result also indicates the existence of scalar hair in the vicinity of the surface.

\subsubsection{Scalaron Profiles}

\begin{figure}[tb]%
  \begin{minipage}[t]{0.5\linewidth}%
    \centering%
    \includegraphics[keepaspectratio, width=0.95\linewidth]{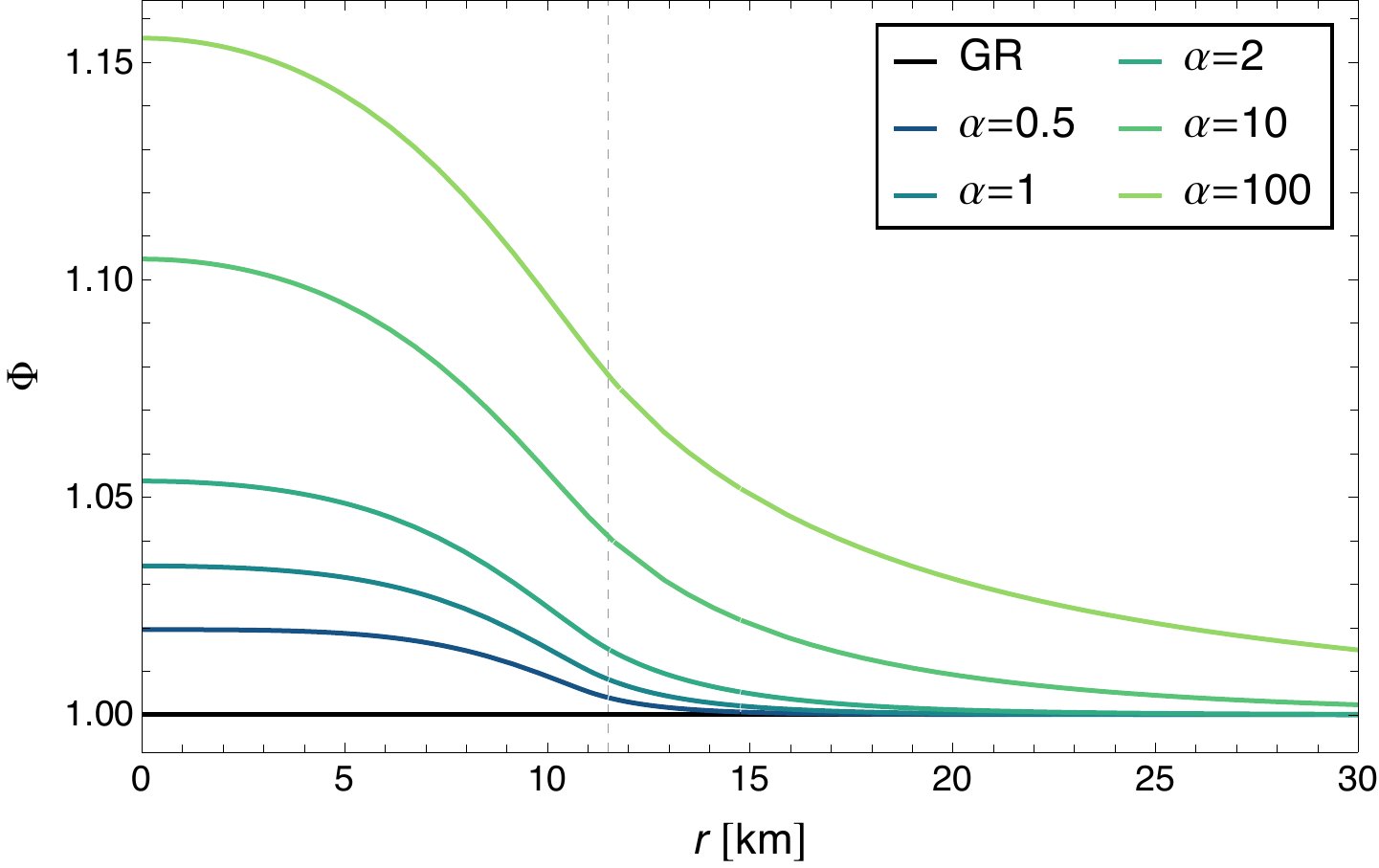}%
    \subcaption{scalaron field distribution around the star.}%
    \label{fig:phi_star}%
  \end{minipage}%
  \begin{minipage}[t]{0.5\linewidth}%
    \centering%
    \includegraphics[keepaspectratio, width=0.95\linewidth]{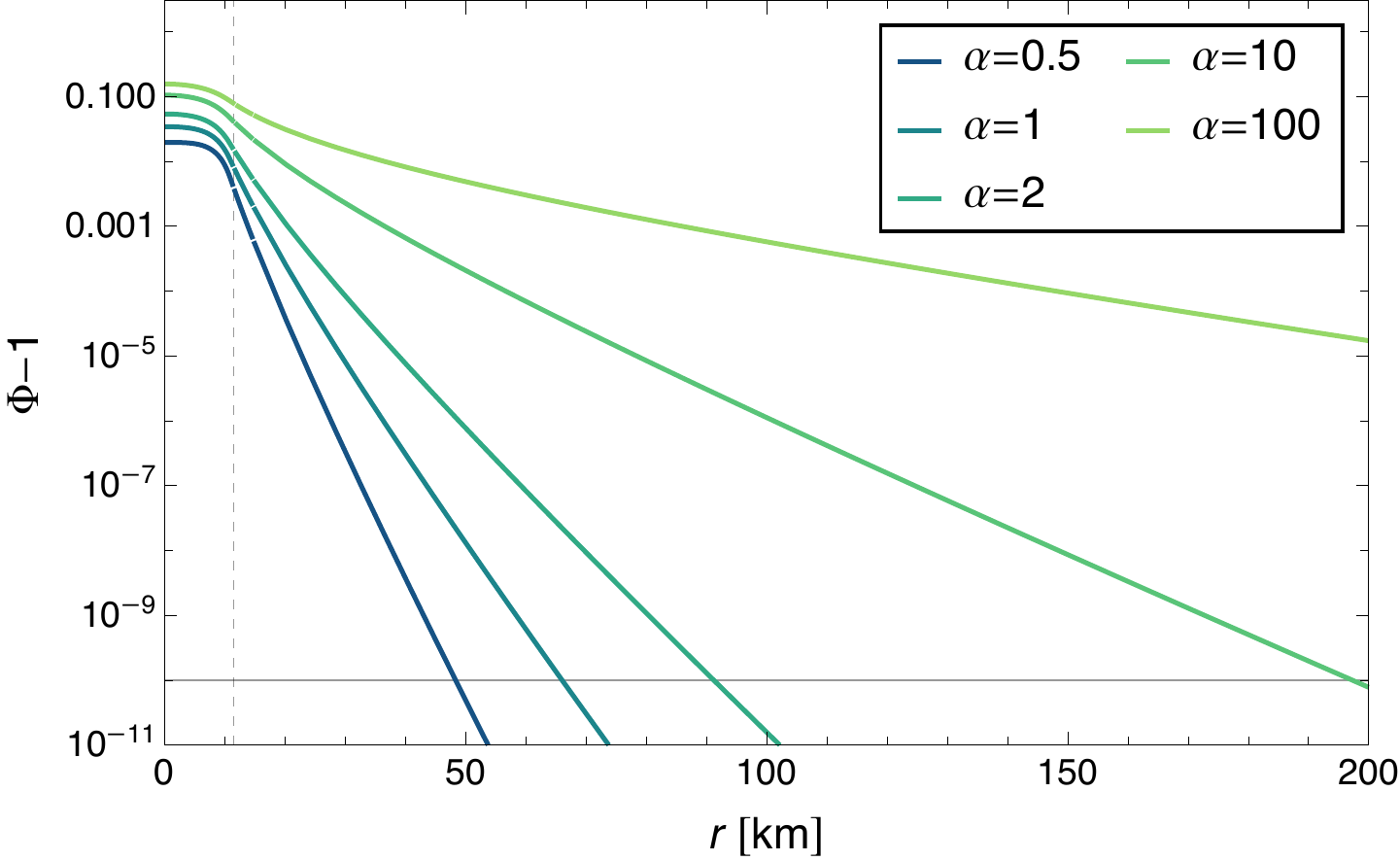}%
    \subcaption{Scalar hair outside the star.}%
    \label{fig:phi-1}%
  \end{minipage}%
  \caption{
The profiles of scalaron \eqref{eq:scalar_BD_R2} for several $\alpha$. 
The values of the central pressure are all $\rho_c = 1.00 \cdot 10^{15}\; \si{g/cm^3}$ using SLy EOS. (a) Profiles of the scalaron in the vicinity of the star. 
The dashed vertical line is the surface radius for GR as a reference value. Nontrivial scalaron concentration gets remarkable as $\alpha$ increases. (b) Profiles of deviations from GR value ($\Phi=1$), i.e., scalar hair outside the star. 
The horizontal solid line represents  $\Phi -1  = 10^{-10}$. 
One can find that the decay of the scalaron becomes mild, and the radius $r_{\Phi}$ defined as $\Phi (r_\Phi)-1  = 10^{-10}$ is prolonged as $\alpha$ increases.
}
  \label{fig:phi}
\end{figure}%

\begin{table}[tbp]
    \centering
    \begin{tabular}{cccc}
         \hline
         $\alpha$ ($r_g$) & Mass $M$ ($M_\odot$) & Radius $r_s$ ($\si{km}$) & Scalarization radius $r_{\Phi}$ ($\si{km}$)\\
         \hline \hline
         0 & 1.930 & 10.880 & -\\
         0.5 & 1.947 & 10.923 & 46.717\\
         1 & 1.958 & 10.967 & 64.077\\
         2 & 1.975 & 11.032 & 88.930\\
         10 & 2.030 & 11.228 & 193.985\\
         \hline
    \end{tabular}
    \caption{The physical quantities for several $\alpha$ with $\rho_c = 1.00 \cdot 10^{15.15}\; \si{g/cm^3}$.}
    \label{tab:alpha_scradius}
\end{table}

Here we investigate the profiles of the scalaron field $\Phi$ defined in Eq.~\eqref{eq:scalar_BD_R2}. 
It should be noted that the GR solution corresponds to $\Phi=1$ by definition everywhere. 
The plots \figref{fig:phi} show the scalar hair profiles inside and outside the star for $p_c \simeq 3.44\cdot10^{-4}\, p_g$ (corresponding to $\rho_c = 1.00 \cdot 10^{15}\; \si{g/cm^3}$) with the SLy EOS \cite{Alford:2004pf}. 
\figref{fig:phi_star} shows that the nontrivial distribution of the scalaron field becomes prominent as long as the value of $\alpha$ increases. 
It can also be noticed that the scalaron field profiles inside the star are not always decreasing monotonically (for small $\alpha$ in particular) inside the star. 

\begin{figure}[tb]%
  \begin{minipage}[t]{0.5\linewidth}%
    \centering%
    \includegraphics[keepaspectratio, width=0.9\linewidth]{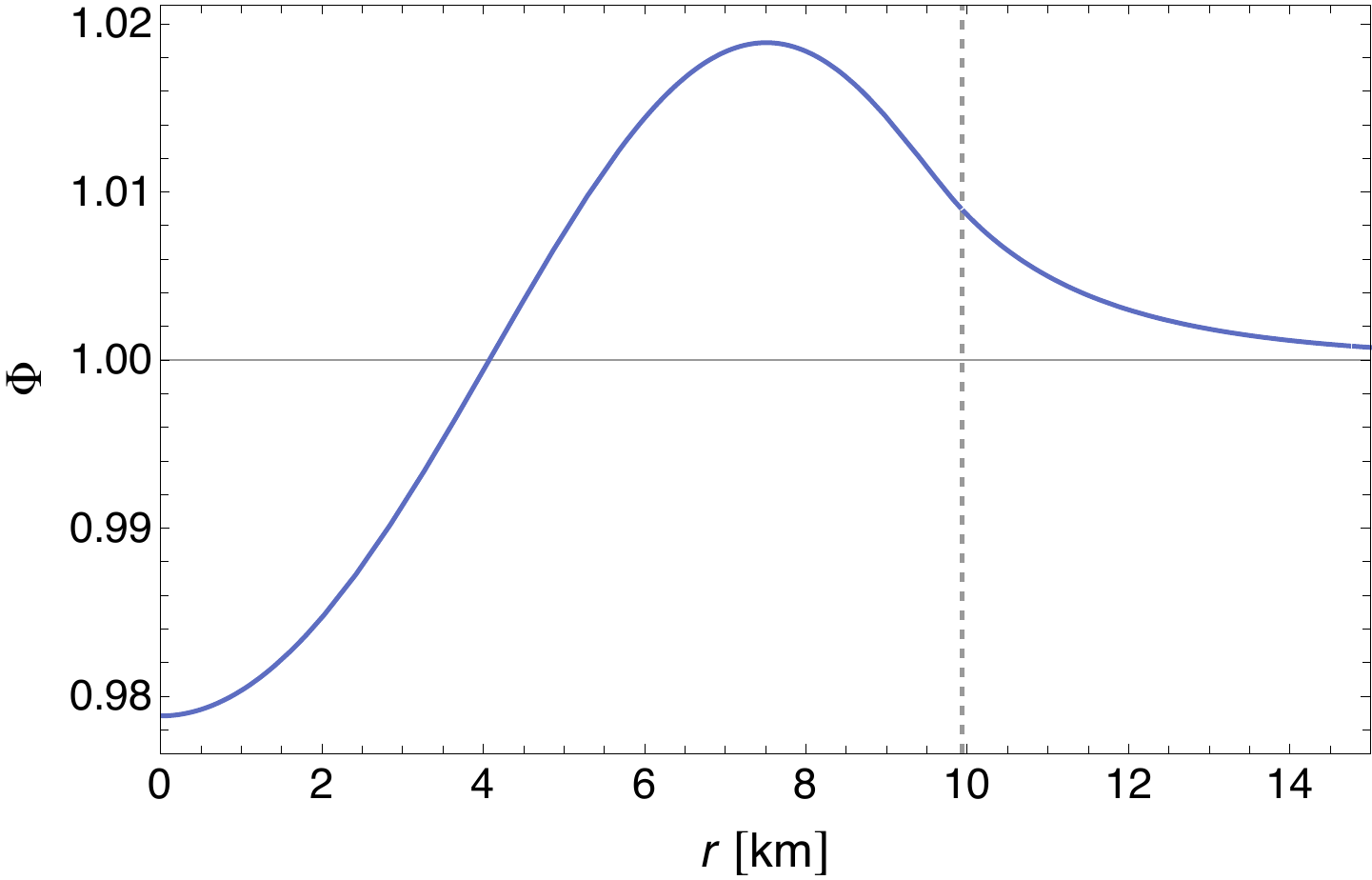}%
    \subcaption{The scalaron profile}%
    \label{fig:phi_star_E15.30}%
  \end{minipage}%
  \begin{minipage}[t]{0.5\linewidth}%
    \centering%
    \includegraphics[keepaspectratio, width=0.9\linewidth]{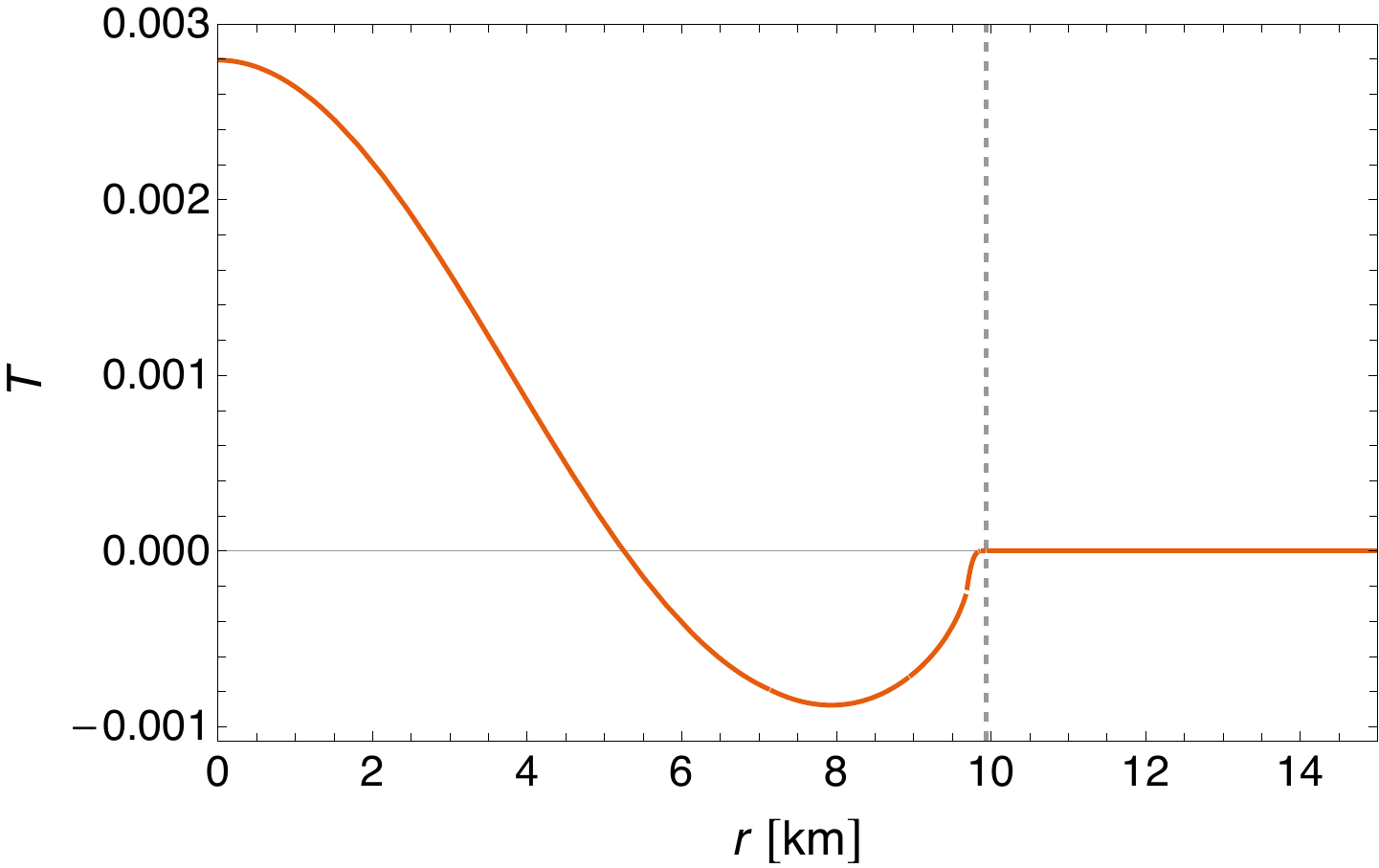}%
    \subcaption{The profile of the trace of energy-momentum $T$}%
    \label{fig:T_E15.30}%
  \end{minipage}%
  \\[5pt]
  \begin{minipage}[t]{\linewidth}%
    \centering%
    \includegraphics[keepaspectratio, width=\linewidth]{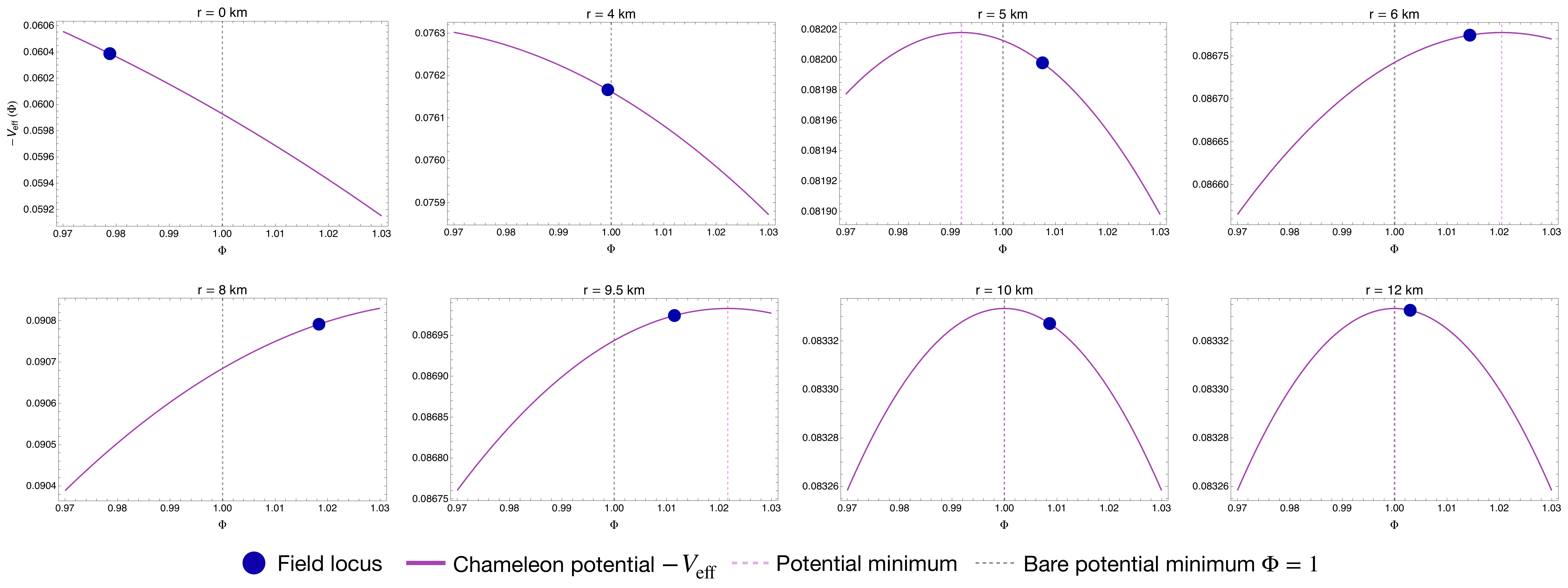}%
    \subcaption{The sign--flipped chameleon potential and the scalaron field value for some radii}%
    \label{fig:inv_pot_anima}%
  \end{minipage}%
  \caption{
The plots for the scalaron and the matter profiles and corresponding chameleon potential with $\alpha=1, \, \rho_c = 1.00 \cdot 10^{15.3}\; \si{g/cm^3}$, and SLy EOS. 
The characteristic behavior of the scalaron field as a function of $r$ in FIG.~(a) can be described by the dynamical motion under the sign--flipped chameleon potential as FIG.~(c). 
Note that the range of the vertical axis in FIG.~(c) is not fixed in order that the property of the scalaron field is not determined by the absolute value of the potential but by its slope.
FIG.~(b) shows the trace of the energy-momentum tensor $T=-(\rho-3p)$. And one finds positive $T$ inside the star, which indicates the existence of stiff matters \cite{Haensel:2007yy}.}
\label{fig:phi_E15.30}
\end{figure}%

In fact, these internal solutions can be understood by the effective behavior of the scalaron field $\Phi$. 
Let us consider the case with $\alpha=1$, $p_c \simeq 3.23\cdot10^{-3}\, p_g$ (corresponding to the $\rho = 1.00 \cdot 10^{15.3}\; \si{g/cm^3}$), and SLy EOS \figref{fig:phi_E15.30}. 
The scalaron field profile, in this case, fluctuates around $\Phi=1$ inside the star and asymptotically coincides with the profile outside, as we find in \figref{fig:phi_star_E15.30}. 
This profile is described by the chameleon potential, which is determined by the trace of the energy-momentum tensor $T$, whose behavior is shown in \figref{fig:T_E15.30}. 
For the scalaron field equation in the $R^2$ gravity
\begin{align}
    \Box \Phi = \pdv{V_{\mathrm{eff}}}{\Phi} \qty(\Phi, T)
    = m_{\Phi}^2 \qty(\Phi - \Phi_{\min}(T))\, ,
\end{align}
we are now considering only radial dependence $r$. 
Taking the signs of the metric into account, solving this equation can be recognized roughly as solving the dynamical (time-dependent) problem under the sign-flipped chameleon potential $-V_{\mathrm{eff}}(\Phi)$ which is now a concave function of $\Phi$. 
The plots in \figref{fig:inv_pot_anima} are the behaviors of $-V_{\mathrm{eff}}(\Phi)$ as a function of the scalaron for several values of the radius. 
We find that the value of $\Phi(r)$ is determined as if the particle slips down this sign--flipped potential $-V_{\mathrm{eff}}(\Phi)$. 
The mentioned fluctuation around $\Phi=1$ is found to stem from the energy-momentum dependence of the effective potential; i.e., the fluctuation of $T$ around $T=0$ causes the fluctuation of the scalaron field. 

\begin{figure}[tb]%
  \begin{minipage}[t]{0.5\linewidth}%
    \centering%
    \includegraphics[keepaspectratio, width=0.95\linewidth]{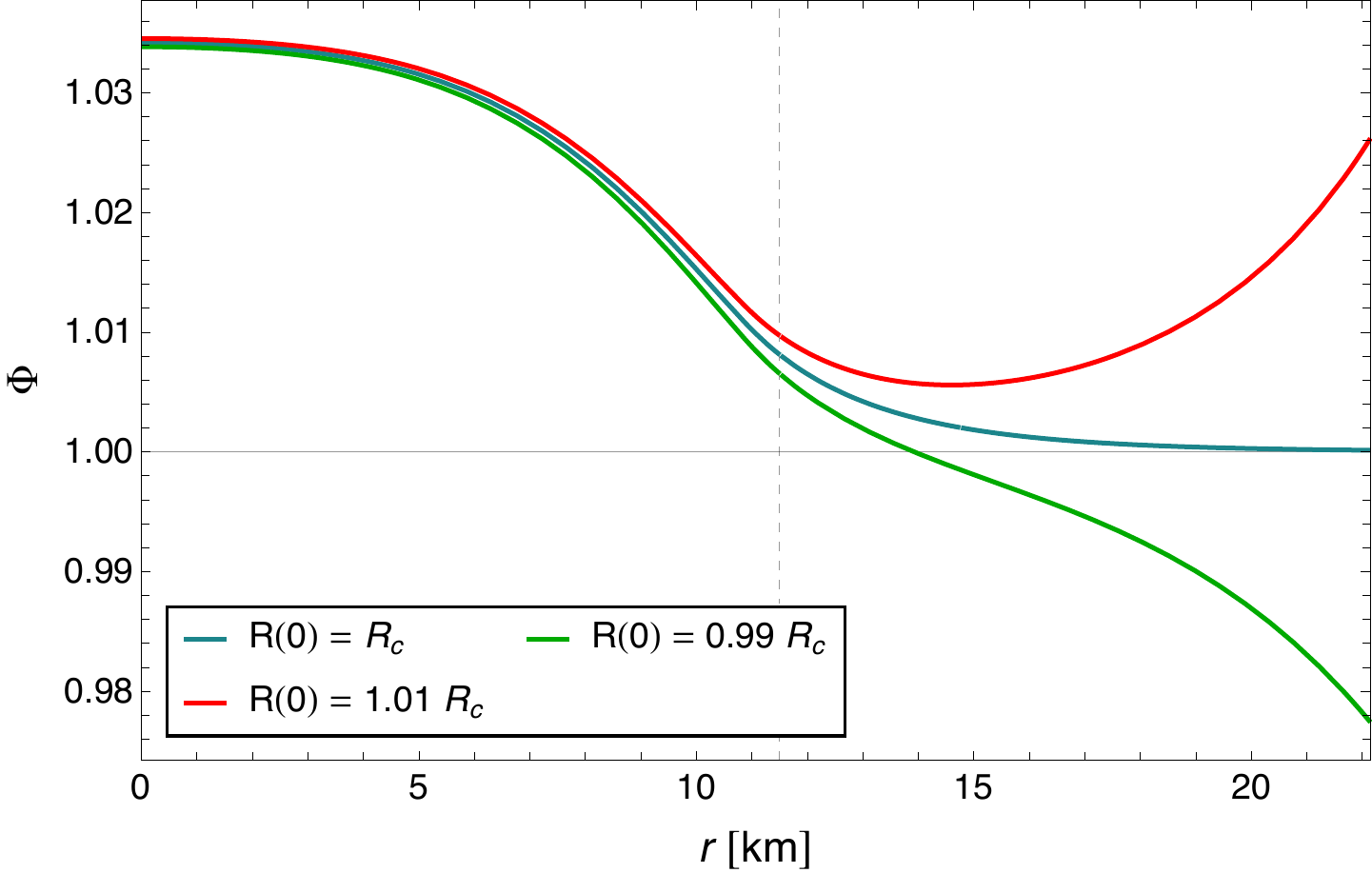}%
    \subcaption{$\alpha=1$}%
    \label{fig:div_1}%
  \end{minipage}%
  \begin{minipage}[t]{0.5\linewidth}%
    \centering%
    \includegraphics[keepaspectratio, width=0.95\linewidth]{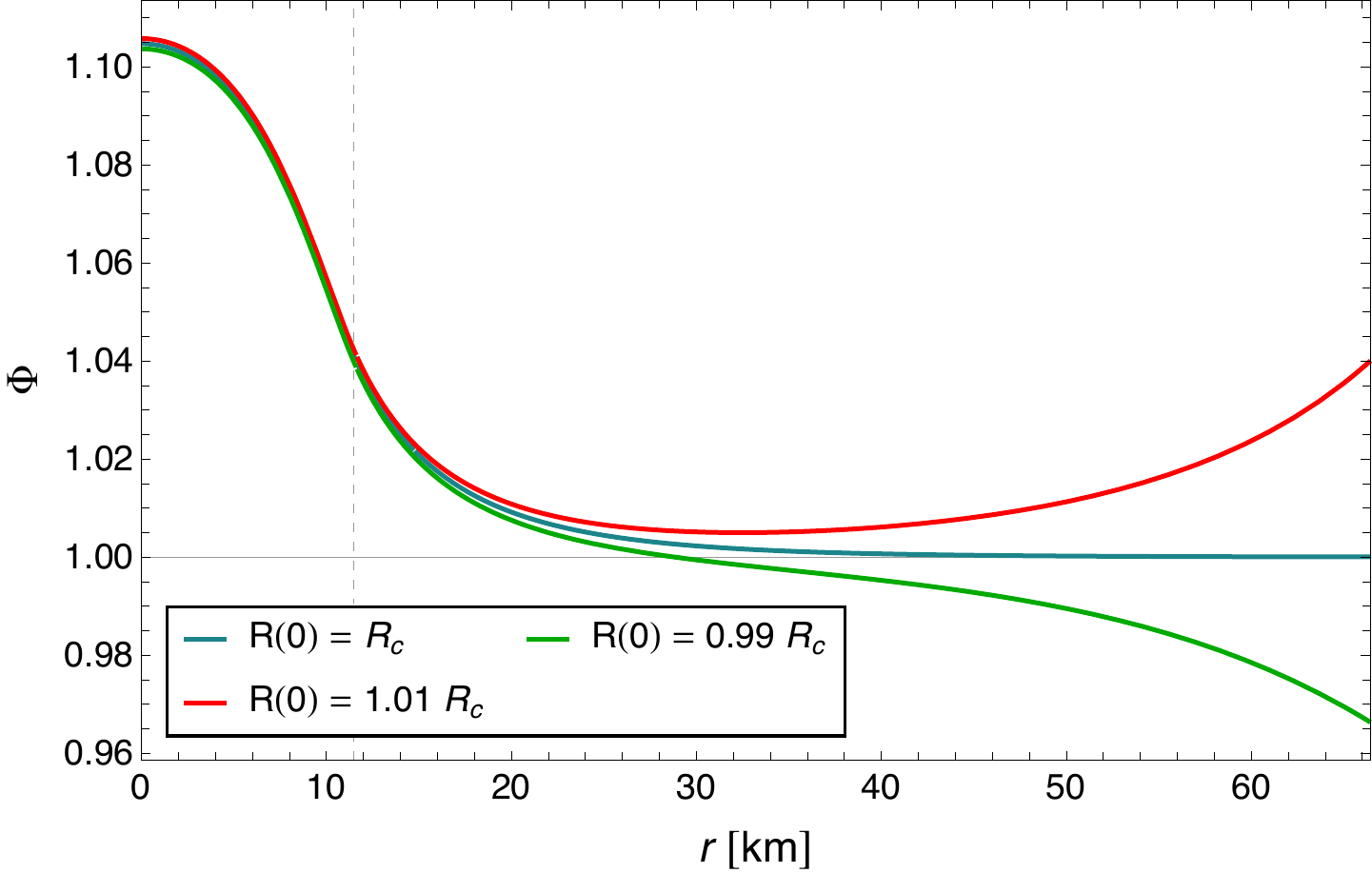}%
    \subcaption{$\alpha=10$}%
    \label{fig:div_2}%
  \end{minipage}%
  \caption{
The plots show the initial-value dependence of the scalaron field distribution. 
For both values of $\alpha$, slight differences cause serious brow-ups of the scalar hair and prevent it from being asymptotically flat. 
Hence such solutions need fine-tuned initial conditions, as also understood from \figref{fig:phi_E15.30}.}
  \label{fig:div}
\end{figure}%
Moreover, the scalaron field finally tends to stay on the top of the concave function $-V_{\mathrm{eff}}(\Phi)$. 
Therefore the asymptotically flat solution is an {\it unstable} branch.
In other words, to find such solutions is to find the proper initial or boundary conditions so that the scalaron field gets {\it the potential maximum} of the $-V_{\mathrm{eff}}(\Phi)$ for large $r$. 

The initial value dependence of the scalaron field distribution is explicitly shown in \figref{fig:div}. 
We can see that a tiny deviation from proper initial values leads to the blowing behavior of the scalar hair. 
The behavior becomes more sensitive for lower $\alpha$. 
This is why the numerical integration is unstable, and we encounter several difficulties in finding solutions. 
This situation was mentioned in some previous works \cite{Resco:2016upv, Kase:2019dqc}. 
It should be noted that this instability is with respect to the radial perturbation, not the dynamical one. 
Thus the validity of the asymptotically flat solution in this gravity depends on the setting one considers.

\figref{fig:phi-1} is the plot for a larger scale. 
It can be noticed that the scalar hair decreases exponentially, and its slope becomes softer as $\alpha$ increases. 
This feature can be confirmed more quantitatively from \tabref{tab:alpha_scradius}. 
This is the summary table of the physical quantities for several values of $\alpha$. 
Here we define the scalarization radius $r_\Phi$, which coincides with a radius of the scalar hair sphere, and $|\Phi (r_\Phi) -1| = 10^{-10}$ numerically. 
\tabref{tab:alpha_scradius} also illustrates the effective gravitational radius of the star in terms of curvature. 
We can find that the differences in the stellar mass $M$ and the stellar radius $r_s$ are $\order{10^{-2}}$ of the values. 
However, the scalarized radii $r_\Phi$ are significantly different for different $\alpha$. 
Moreover, the ratio of the scalarized radii is found to be roughly proportional to the ratio of the Compton length of the scalaron field $\ell = \sqrt{6\alpha}$: for instance, $r_{\Phi}(\alpha=10)/r_{\Phi}(\alpha=1) \simeq \sqrt{10}$. 
These results imply the existence of scalar hairs with the Compton length of the scalaron field in the $R^2$ gravity, which is characterized by the parameter for the $R^2$ correction term.

\subsubsection{Mass-Radius Relation}

\begin{figure}[tb]%
  \begin{minipage}[t]{0.5\linewidth}%
    \centering%
    \includegraphics[keepaspectratio, width=0.95\linewidth]{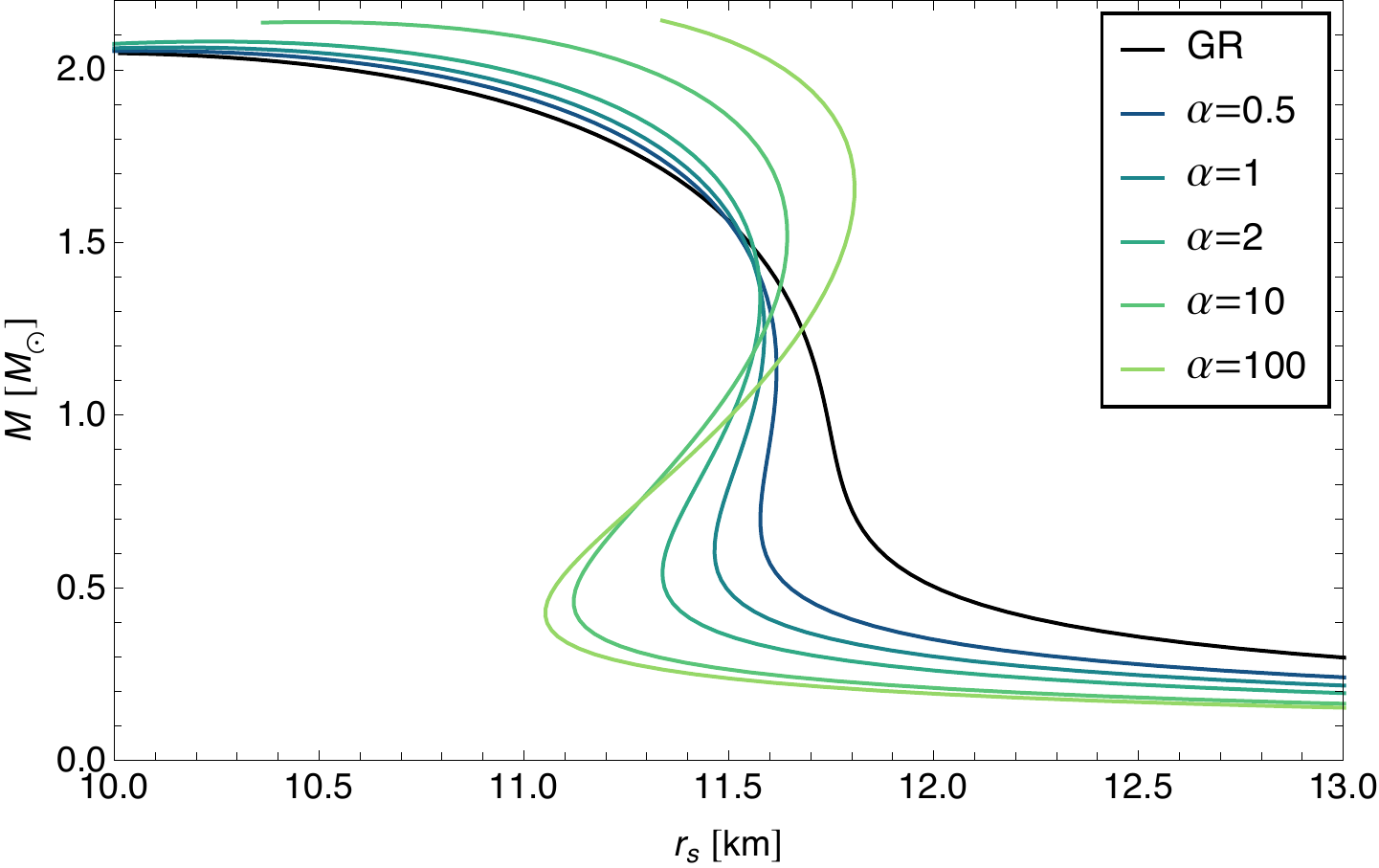}%
    \subcaption{$M$--$r_s$ relation with SLy EOS}%
    \label{fig:MR_SLy}%
  \end{minipage}%
  \begin{minipage}[t]{0.5\linewidth}%
    \centering%
    \includegraphics[keepaspectratio, width=0.95\linewidth]{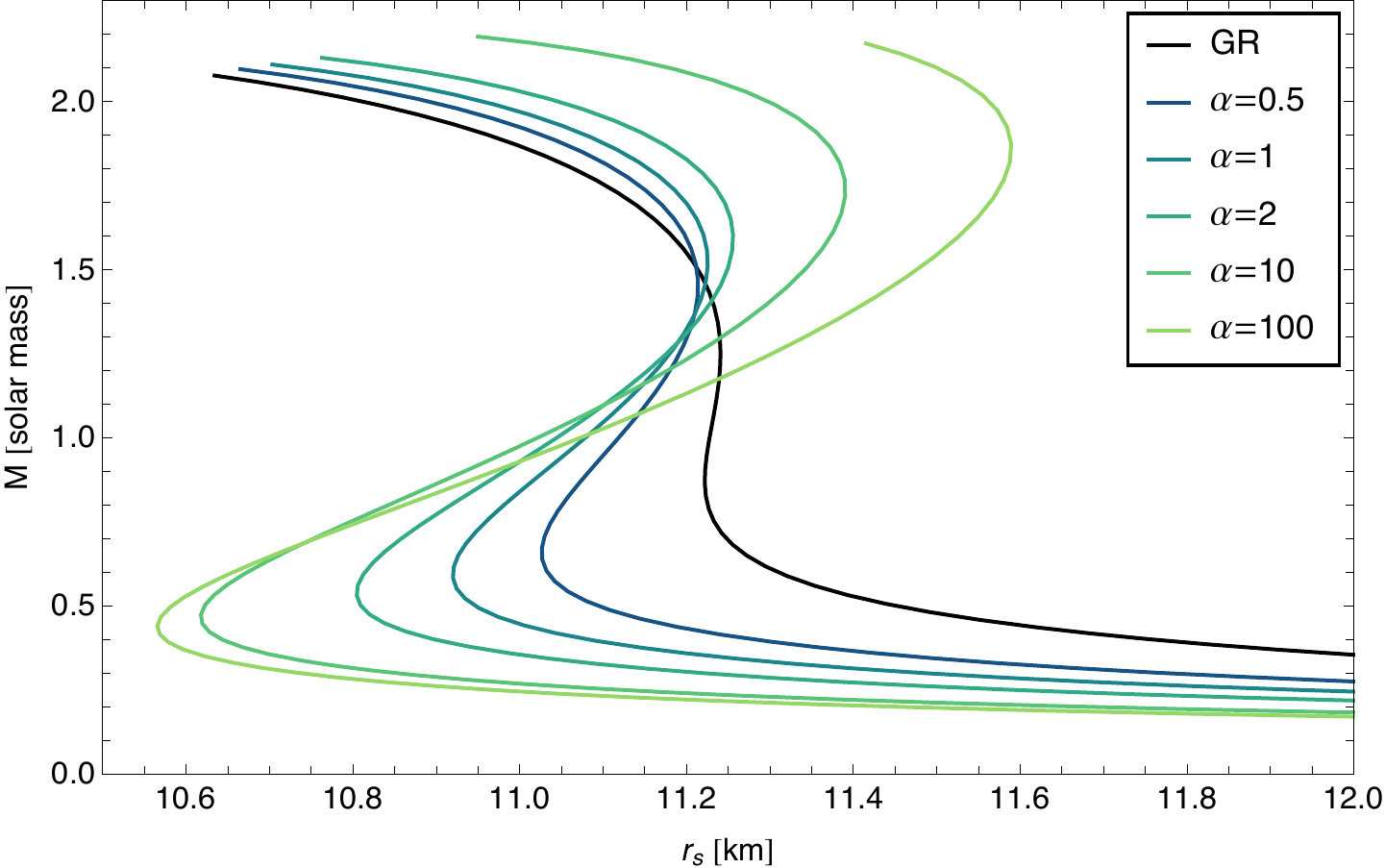}%
    \subcaption{$M$--$r_s$ relation with APR4 EOS}%
    \label{fig:MR_APR}%
  \end{minipage}\\[5pt]%
  \begin{minipage}[t]{0.5\linewidth}%
    \centering%
    \includegraphics[keepaspectratio, width=0.95\linewidth]{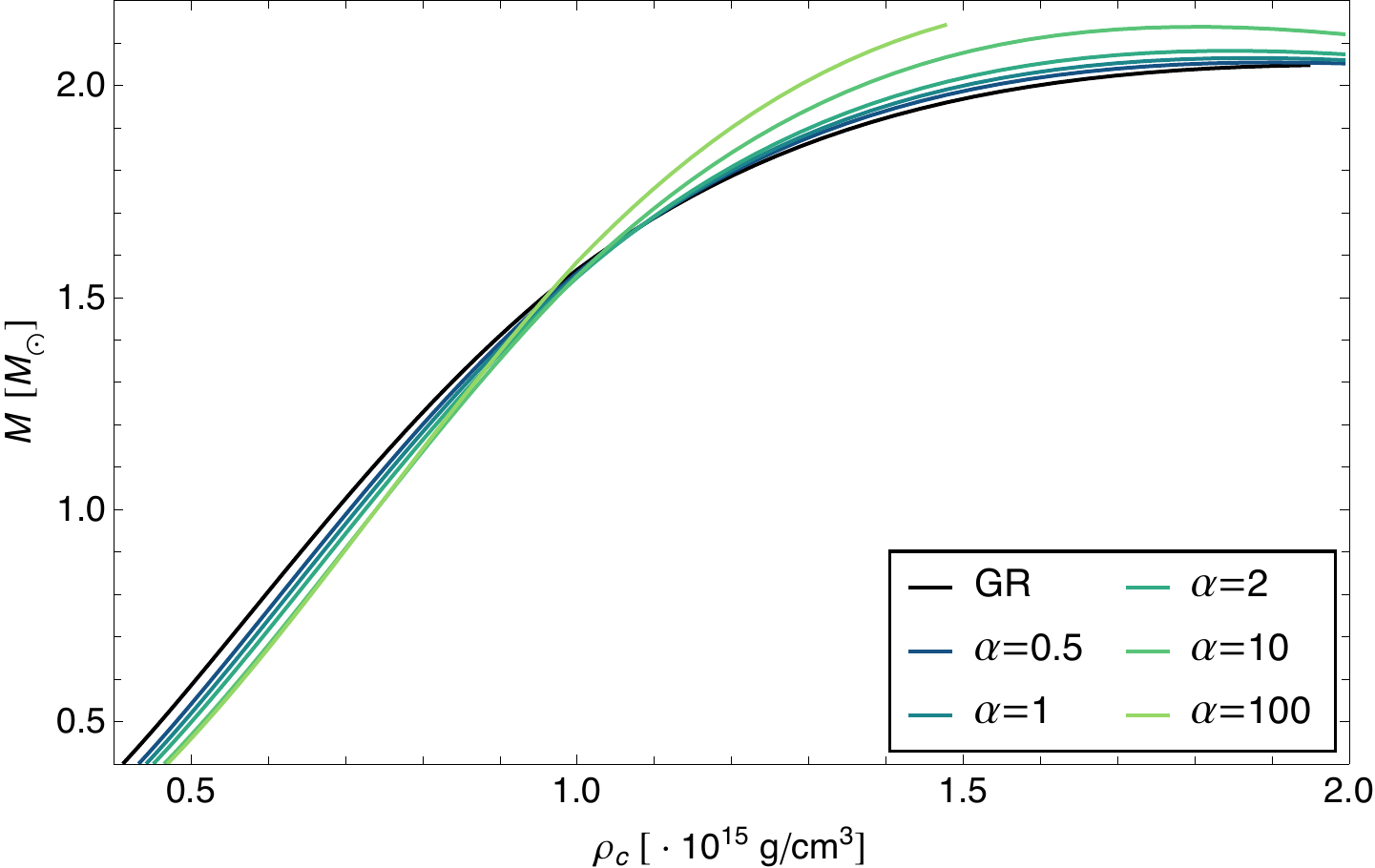}%
    \subcaption{$M$--$\rho_c$ relation with SLy EOS}%
    \label{fig:Mrho_SLy}%
  \end{minipage}%
  \begin{minipage}[t]{0.5\linewidth}%
    \centering%
    \includegraphics[keepaspectratio, width=0.95\linewidth]{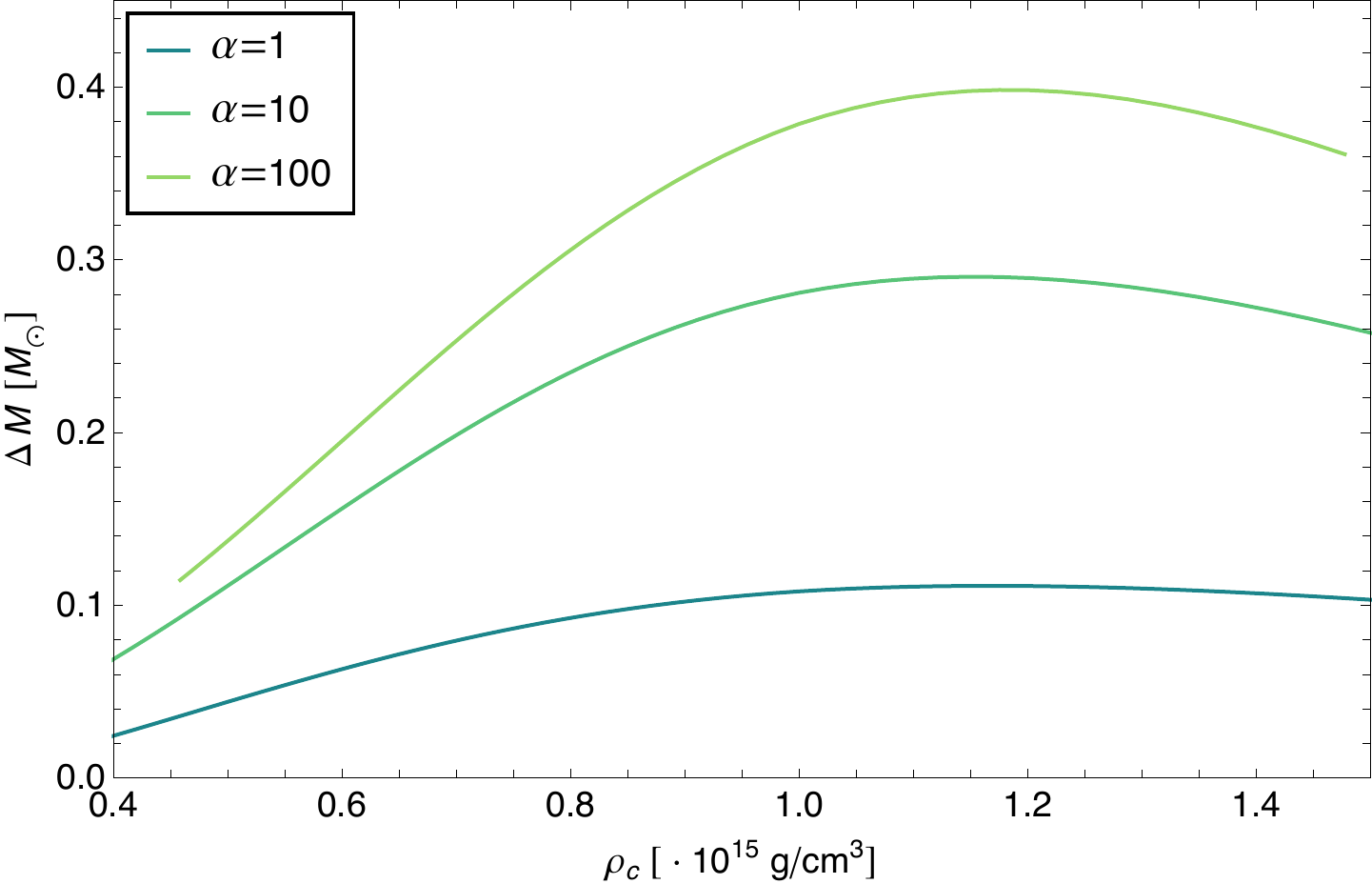}%
    \subcaption{Differences between $M$ and $m(r_s)$}%
    \label{fig:mass_diff}%
  \end{minipage}
  \caption{
(a), (b) The mass-radius relation with two EOSs for several values of $\alpha$. 
The mass is defined as the Schwarzschild mass $M$ at the endpoint $r_e$, and the radius is the surface radius $r_s$. 
The curves seem to rotate clockwise as $\alpha$ increases regardless of the EOSs. 
(c) The mass as the function of the central rest mass energy density $\rho_c$. The relation between $\alpha$ and $M$ changes with respect to $\rho_c$. 
(d) The difference $\Delta M$ between the stellar mass $M$ and the effective Schwarzschild mass on the surface $m(r_s)$. 
The difference shows the mass contribution of the scalar hair outside the star to the mass and increases as we increase $\alpha$.
}
  \label{fig:MR}
\end{figure}%

Repeating the calculation procedure for the desired range of the central pressure $p_c$, we found the mass-radius ($M$--$r_s$) relation for each EOS and for several values of $\alpha$. 
The results are shown in \figref{fig:MR_SLy} and \figref{fig:MR_APR} for two different EOSs SLy \cite{Alford:2004pf} and APR4 \cite{Douchin:2001sv}.
In both of the two plots, the $M$-$r_s$ curves are distorted as if they {\it rotate clockwise} in response to $\alpha$ around some fixed points. 
The maximum masses become heavier as $\alpha$ increases. 
These results are consistent with those in the existing works such as \cite{Yazadjiev:2014cza, Astashenok:2018iav}. 
Therefore, these behaviors are qualitatively independent of EOSs but strongly depend on the correction parameter $\alpha$. 
Obtained results are supposed to stem from the modification of the gravity theory, that is, the additional scalar DOF.

To clarify the effect of the scalaron field on the stellar mass, we plot the relationship between the stellar mass $M$ and the center rest mass energy $\rho_c$ with SLy EOS in \figref{fig:Mrho_SLy}. 
We can see that $M$ increases as $\alpha$ does for the high $\rho_c$ region. 
After some turning point (around $\rho_c \simeq 1.1 \cdot 10^{15}\; \si{g/cm^3}$), however, $M$ decreases as $\alpha$ increases in the lower $\rho_c$ region. 
Hence the scalaron field contributes to the stellar mass as {\it negative energy} in effect. 
It does not always have a positive influence on $M$, and this behavior corresponds to the {\it rotation} of the mass-radius relation we just mentioned.

\begin{figure}[tb]%
  \begin{minipage}[t]{0.5\linewidth}%
    \centering%
    \includegraphics[keepaspectratio, width=0.95\linewidth]{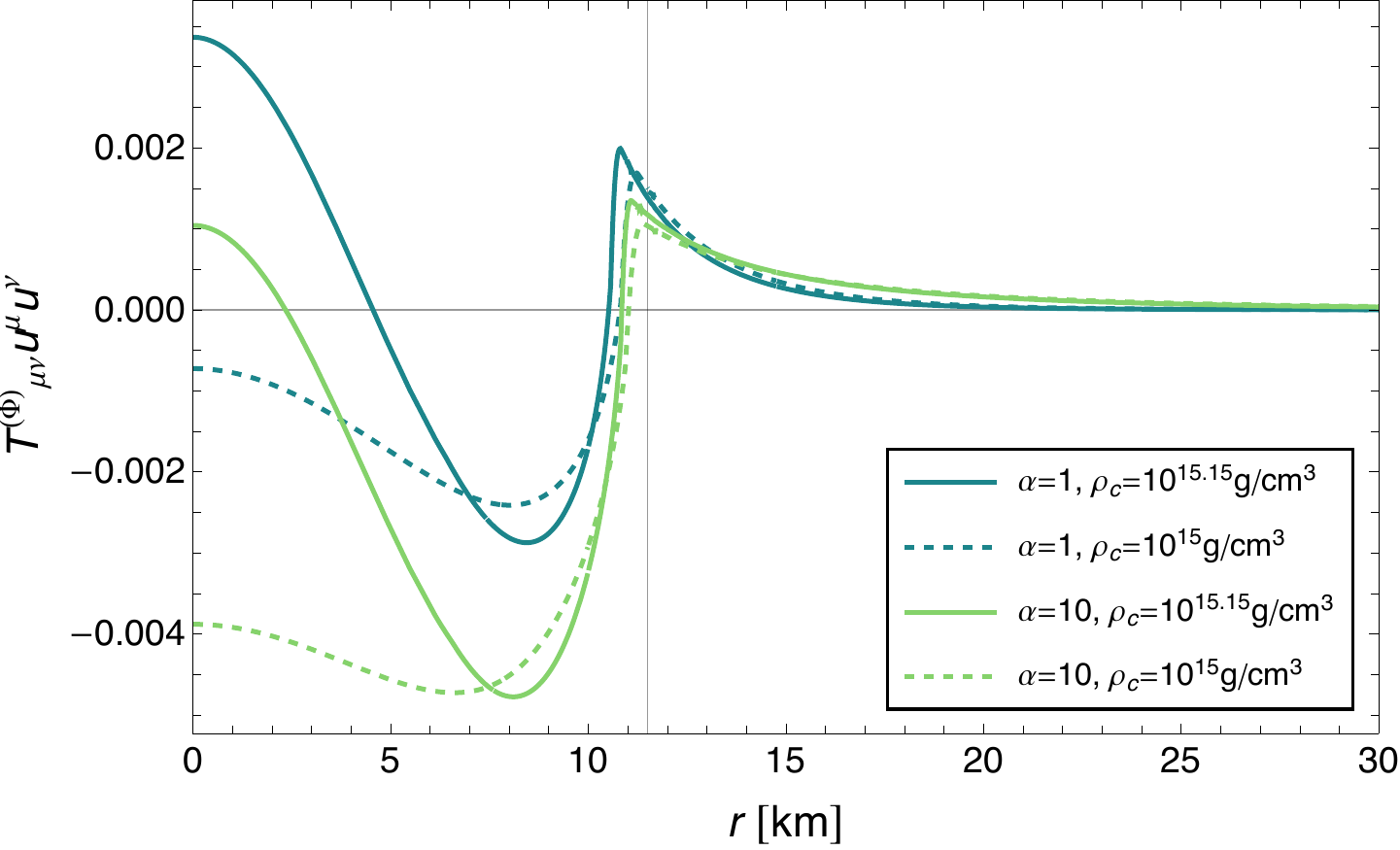}%
    \subcaption{$T^{(\Phi)}_{\mu\nu} u^{\mu} u^{\nu}$}%
    \label{fig:EMcond_timelike}%
  \end{minipage}%
  \begin{minipage}[t]{0.5\linewidth}%
    \centering%
    \includegraphics[keepaspectratio, width=0.95\linewidth]{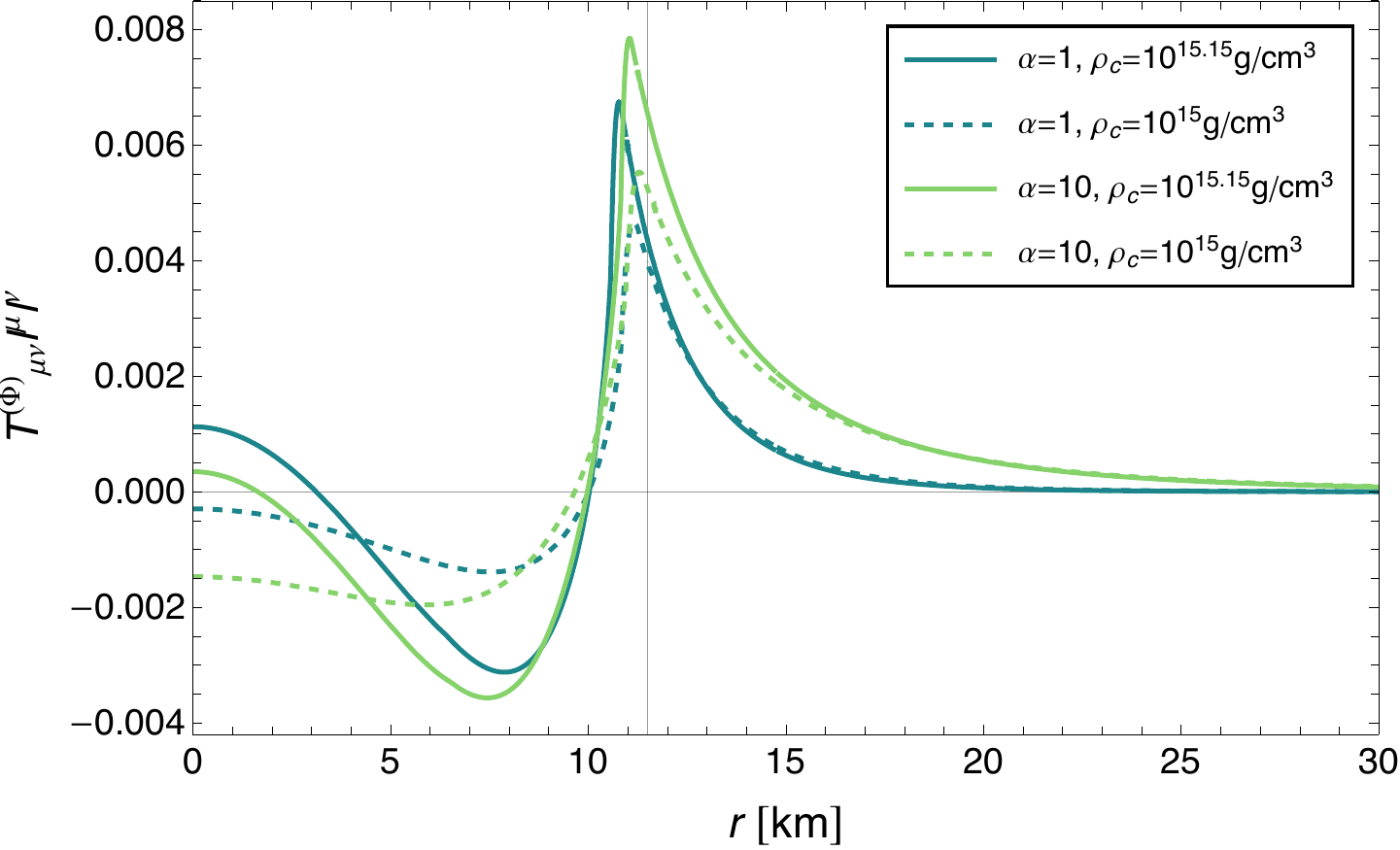}%
    \subcaption{$T^{(\Phi)}_{\mu\nu} l^{\mu} l^{\nu}$}%
    \label{fig:EMcond_null}%
  \end{minipage}%
  \caption{
The effective energy conditions for the scalaron field as functions of the radius $r$. 
Inside the star, the conditions can be partially broken, and the scalaron behaves as a quintessence field in this broken region. 
These features become significant for lower $\rho_c$. 
The quintessential scalaron field effectively gives negative energy densities (which corresponds to (a)). 
It is one of the reasons for the nonmonotonic relation of $M$ and $\rho_c$ in \figref{fig:Mrho_SLy}.
}
  \label{fig:EMcond}
\end{figure}%

The reason can be understood from the effective energy condition of the scalaron field $\Phi$. 
The effective energy-momentum tensor $T^{(\Phi)}_{\mu\nu}$ of the scalaron field $\Phi$ can be constructed as Eq.~\eqref{eq:EMtensor_scalar}. 
Using this expression, we can evaluate the effective energy condition of the scalaron field using $T^{(\Phi)}_{\mu\nu} t^{\mu} t^{\nu}$ and $T^{(\Phi)}_{\mu\nu} \ell^{\mu} \ell^{\nu}$ with arbitrary timelike-- and null vector $t^{\mu}$ and $\ell^{\mu}$. 

\figref{fig:EMcond} show the value of these conditions for some values of $\rho_c$ and $\alpha$ respectively~\footnote{To compare among the different $\alpha$, we normalized $T^{(\Phi)}_{\mu\nu}$ by $\Phi(r)$ as noticed from \eqref{eq:EOM_metric_BD} in \figref{fig:EMcond}.}. 
Here we use the four--velocity of the fluid $u^{\mu}$ and one null vector $l^{\mu} = \{\e^{\lambda-\nu}, 1, 0, 0\}$ for evaluation. 
The trends differ between the external- and the internal region of the star. 
For all parameter settings, both of the energy conditions indicate positive values outside the star. 
Thus it is expected that $\Phi$ behaves as an ordinary baryonic matter and that the scalar hair weights the star. 
This fact also can be seen from \figref{fig:mass_diff}. 
It shows the difference between the effective Schwarzschild mass on the surface $r=r_s$ and on the spacial infinity $r\rightarrow \infty$; 
\begin{align}
    \Delta M = M - m(r_s) \qq{where} m(r) \equiv \frac{r}{2} \qty(1-\e^{-2\lambda(r)})\,,
\end{align}
as function of $\rho_c$. 
Thus it evaluates the whole energy contribution from the scalar hair. We should notice that $\Delta M$ is always positive for all parameter settings. 
Therefore scalar hair outside the star always makes the star heavier, and this tendency becomes more drastic as $\alpha$ increases.

On the other hand, the situation is different inside the star. 
As we can see in \figref{fig:EMcond}, both conditions can take negative values. 
In particular, they become negative almost all regions except for the vicinity of the surface for lower $\rho_c$ settings. 
Thus the scalaron field $\Phi$ can break the all of energy conditions and behaves as the quintessence field. 
As a result, it can decrease the entire stellar mass $M$, especially for lower $\rho_c$ solutions. 
This quintessential feature itself is not surprising in $F(R)$ gravity theory as mentioned in \cite{Capozziello:2002rd}.

In summary, the scalaron field works differently for the external and internal regions; it acts as a baryonic matter with positive contributions and as a (partially) quintessential matter with negative contributions to the mass, respectively. 
Also, the coupling between the matter and the Einstein gravity field (i.e., the metric $g_{\mu\nu}$) is effectively masked by the scalaron, as noticed from Eq.~\eqref{eq:EOM_metric_BD}. 
The stellar mass $M$ is affected by all of these contributions from the scalaron. 
As a result, the star becomes more massive or less massive as \figref{fig:Mrho_SLy} depending on the theory modification parameter $\alpha$ and the central condition such as $\rho_c$.


\section{Conclusion \label{sec:conclusion}}

In this paper, we investigated the TOV configuration of compact stars in the $R^2$ gravity theory \eqref{eq:R^2}. 
To construct the asymptotically flat solution, we utilized the chameleon potential in Eq.~\eqref{eq:cham_pot_R^2_Jordan} to analyze the spatial distribution of the additional scalar DOF. 
Considering the asymptotic behavior of the scalaron field, we clarified that the outer asymptotic geometry has exponentially decreasing curvature corresponding to the scalar hair~\eqref{eq:phi_boundary}. 
In addition, we specified the two-boundaries shooting method as a suitable choice for the numerical integration, in which we can avoid exponentially diverging numerical instabilities coming from this scalaron behavior.

Based on the above forecasts, we performed the numerical integration of the modified TOV equations Eqs.~\eqref{eq:continuity_TOV}--\eqref{eq:R''_R2}. 
We found that the geometry and the matter distribution nontrivially deviate from the GR solution around the star as \figref{fig:geo_SLy} and that the neuron stars have expected scalar hairs which depend on the effective mass of the scalaron as \figref{fig:phi}. 
For the internal region of the star, we explicitly showed in \figref{fig:phi_E15.30} that the qualitative trend of the scalaron is determined by the chameleon potential $V_{\mathrm{eff}} (\Phi)$ as if the dynamical problem under the sign-flipped concave-upward potential $-V_{\mathrm{eff}} (\Phi)$. 
This shape of the potential also explains the numerical difficulties in finding the proper solutions. 
The scalaron field influences the stellar mass examined in \figref{fig:MR} with its energy conditions. 
We revealed in \figref{fig:EMcond} that the external scalar hair gives positive mass, while the internal quintessential scalaron can give negative contributions. 
All these effects collectively influence the stellar mass and make it heavier for higher $\rho_c$ solutions and lighter for lower $\rho_c$ ones.

As seen in this paper, we uncovered several ambiguous points remaining in the previous works. 
The asymptotically flat external geometry of compact stars in the $R^2$ gravity (i.e., the existence of the scalar hair) had been a controversial issue. 
Our results for the outer geometry support Refs.~\cite{Yazadjiev:2014cza, Astashenok:2017dpo, Astashenok:2018iav} which showed neutron stars have scalar hair, while opposing Refs.~\cite{Ganguly:2013taa, Feng:2017hje} which claimed the neutron stars have no hair in this gravity~\footnote{Although we also disagree with Refs.~\cite{Resco:2016upv, Feola:2019zqg} for the physical viability of negative $\alpha$, our results do not directly disapprove their numerical results because of the different region for parameter $\alpha$.}. 
This fact raises doubt on the existence of the no-hair theorem for horizon-less compact stars in the $R^2$ gravity discussed in Refs.~\cite{Whitt:1984pd, Mignemi:1991wa, Nzioki:2009av}. 
We also shed light on the inner profile and the energy conditions of the scalaron field, which fulfill the missing parts in the previous works. 
It was clarified that the sign of $\alpha$ and the distribution of the scalaron in varying chameleon potential reflects aforementioned numerical difficulties, which the previous works revealed individually~\footnote{In Ref.~\cite{Kase:2019dqc}, the authors chose the Schwarzschild solution as the external geometry due to this numerical difficulty. 
We revealed that it is not necessary to take such a strict condition if one chooses the proper integration method as we showed.}.
In addition, we showed that the effective behavior of the scalaron field implied by the energy conditions gives a clearer reason for the shape of the mass-radius relation, which had been widely confirmed.

The main point of this work is to fully utilize the effective behavior of the scalaron field, which is the main difference from the configuration in GR. 
For the outside star, the brief analysis of the nature of the scalaron field enables us to figure out the asymptotic solutions and even the suitable integration method before actual numerical calculation. 
The estimation of the asymptotic behavior of this additional DOF is essential to comprehend the situation appropriately. 
Also, the investigation of the scalaron field provides us with a natural understanding of its influences on the internal structure and the mass of compact stars. 
In this work, we discussed how the inner scalaron field (or geometry) distribution is determined in terms of the chameleon field and showed that this scalaron affects the stellar mass in a nonmonotonic way. 
The insight into the internal property of the additional DOF would help investigate the internal structure of the star and discuss the observational implication of the modification of the gravity theory.

As we have seen, the existence of the unscreened massive scalaron causes notable effects, such as exponentially decaying scalar hair, possible restoration of the uniqueness of asymptotically Schwarzschild solution, and its influences on the internal structure of the compact star.
This situation probably differs in the other models of the $F(R)$ gravity whose scalaron has an efficient screening mechanism.
For instance, one can consider viable DE models of $F(R)$ gravity.
In these models, the screening mechanism is essential to guarantee compatibility with results in the local gravitational experiments,
and thus the scalaron has varying mass depending on the presence of matters, unlike the $R^2$ model.
One can expect that the internal screened scalaron would be less influential because $F(R)$ gravity tends to approximate GR.
However, the unscreened scalaron outside the compact star could show prolonged external scalar hairs because it should be a light scalar field to account for DE in vacuum.
We will address the reexamination of the modified TOV equation and reveal how the additional DOF works for the other models of $F(R)$ gravity theory in our future works.

It would also be interesting to investigate the relations between the internal scalar DOF and measurements other than the mass-radius relation.
The scalaron field may cause significant effects on the density profile of the neutron star in other types of the $F(R)$ gravity, which can trigger the rapid cooling as discussed in the scalar-tensor theory \cite{Dohi:2020bfs}.
In addition to the density profile, the additional scalaron field can interact with nucleons, neutrinos, and photons of the neutron stars, which also affects the cooling process.
Similar to constraints on new particles captured in neutron stars \cite{Sedrakian:2015krq, Kouvaris:2010vv, Bhat:2019tnz}, we could constrain the scalaron field as a new particle in the $F(R)$ gravity.

\begin{acknowledgements}
We thank S. Tsujikawa, C.-M. Yoo,  K. Izumi, and T. Shiromizu for their helpful comments. The author (KN) would like to take this opportunity to thank the ``Nagoya University Interdisciplinary Frontier Fellowship'' supported by Nagoya University and JST, the establishment of university fellowships towards the creation of science technology innovation, Grant Number JPMJFS2120.
TK is supported by National Key R\&D Program of China (2021YFA0718500) and by Grant-in-Aid of Hubei Province Natural Science Foundation (2022CFB817).

\end{acknowledgements}

\appendix

\section{Einstein Frame Description \label{sec:EinsteinFrame}}

Although the physical situation is apparent in the Jordan frame, the discussion on the DOF of the gravitational field is not. 
We can clarify it by the conformal transformation of the metric:
\begin{align}
    g_{\mu\nu} \rightarrow&\,  
    \Tilde{g}_{\mu\nu}
    = \Phi \, g_{\mu\nu} 
    \equiv \e^{\sqrt{\frac{2}{3}}\kappa \Tilde{\Phi}} \, g_{\mu\nu}\, ,
    \nn
    V(\Phi) \rightarrow&\, 
    U(\Tilde{\Phi}) 
    = \frac{1}{2\kappa^2}\e^{-2\sqrt{\frac{2}{3}}\kappa \Tilde{\Phi}}
    V(\Phi)\, .
\end{align}
The action Eq.~\eqref{eq:action_BD} is rewritten as
\begin{align}
    S_G = 
    \int d^{4} x \sqrt{-\Tilde{g}}
    \qty[\frac{\Tilde{R}}{2 \kappa^{2}} 
    - \frac{1}{2} \partial^{\alpha} \Tilde{\Phi} \partial_{\alpha} \Tilde{\Phi}
    -U(\Tilde{\Phi})]
    \, .
    \label{eq:action_Einstein}
\end{align}
This action comprises the Einstein-action and minimally-coupled canonical scalaron field with potential $U(\Tilde{\Phi})$. 
We can find that the whole DOF of the gravitational field is $2+1$. 
In this way, we can check that mathematically equivalent action Eq.~\eqref{eq:action_Jordan} have the same number of DOF.
This frame is often called the Einstein frame in contrast with the Jordan frame with the action Eq.~\eqref{eq:action_Jordan}. 

The field equations for $\Tilde{g}_{\mu\nu},\, \Tilde{\Phi}$ are
\begin{align}
    &\Tilde{R}_{\mu\nu} - \frac{1}{2} \Tilde{R} \Tilde{g}_{\mu\nu}
    = \kappa^2 \qty(\Tilde{T}_{\mu\nu}+\Tilde{T}^{\Tilde{\Phi}}_{\mu\nu}),
    \nn
    &\qty(
        \Tilde{T}^{\Tilde{\Phi}}_{\mu\nu}
        =\partial_{\mu} \Tilde{\Phi} \partial_{\nu} \Tilde{\Phi}
        - \Tilde{g}_{\mu\nu}
        \qty[
        \frac{1}{2} \partial^{\alpha} \Tilde{\Phi} \partial_{\alpha} \Tilde{\Phi}
        -U(\Tilde{\Phi}) 
        ]
    ) \\
    &\Tilde{\square} \Tilde{\Phi}
    =U^{\prime}(\Tilde{\Phi})
    +\frac{\kappa}{\sqrt{6}} \:  \Tilde{T} \, ,
\end{align}
where
\begin{align}
    \Tilde{T}_{\mu \nu} \equiv \frac{2}{\sqrt{-\Tilde{g}}} \frac{\delta S_M}{\delta \Tilde{g}^{\mu \nu}}
    =\e^{-\sqrt{\frac{2}{3}}\kappa\Tilde{\phi}} \: T_{\mu\nu} \,,
\end{align}
and the tilde denotes the quantities considered in Einstein frame.

One can also discuss the chameleon mechanism in the Einstein frame. 
The chameleon potential $U_{\mathrm{eff}} (\Tilde{\Phi}, \Tilde{T})$ in this frame is defined as
\begin{align}
    \Tilde{\square} \Tilde{\Phi}
    =&\, U^{\prime}(\Tilde{\Phi})
    +\frac{\kappa}{\sqrt{6}} \:  \Tilde{T} \,
    = \frac{1}{\sqrt{6}\kappa}
    \qty[\frac{2F(R(\Tilde{\Phi}))-R(\Tilde{\Phi})F_R(R(\Tilde{\Phi}))+ \kappa^2 T}{F_R^2 (R(\Tilde{\Phi}))} ] \nn
    \equiv &\, \pdv{U_{\mathrm{eff}}}{\Tilde{\Phi}} \qty(\Tilde{\Phi}, \Tilde{T})\, ,
\end{align}
(Note that $T$ is the Jordan frame quantity.). 
The chameleon mass of the scalaron field in the Einstein frame is
\begin{align}
    m_{\Tilde{\Phi}}^2 
    \equiv& \eval{\pdv[2]{U_{\mathrm{eff}}}{\Tilde{\Phi}}}_{\Tilde{\Phi}=\Tilde{\Phi}_{\min}} \nn
    =&
    \frac{1}{3 F_R\left(R_{\min }\right)}\left(\frac{F_R\left(R_{\min }\right)}{F_{R R}\left(R_{\min }\right)}-R_{\min }\right)
    =\frac{1}{F_R\left(R_{\min }\right)} m_{\Phi}^2 \, ,
\end{align}
where $R_{\min} = R(\Tilde{\Phi}_{\min})$ which is realized at potential minimum. 
The dependence on the energy-momentum comes into the chameleon mass via $R_{\min}$ implicitly. 

\begin{figure}[tb]%
  \begin{minipage}[t]{0.5\linewidth}%
    \centering%
    \includegraphics[keepaspectratio, width=0.9\linewidth]{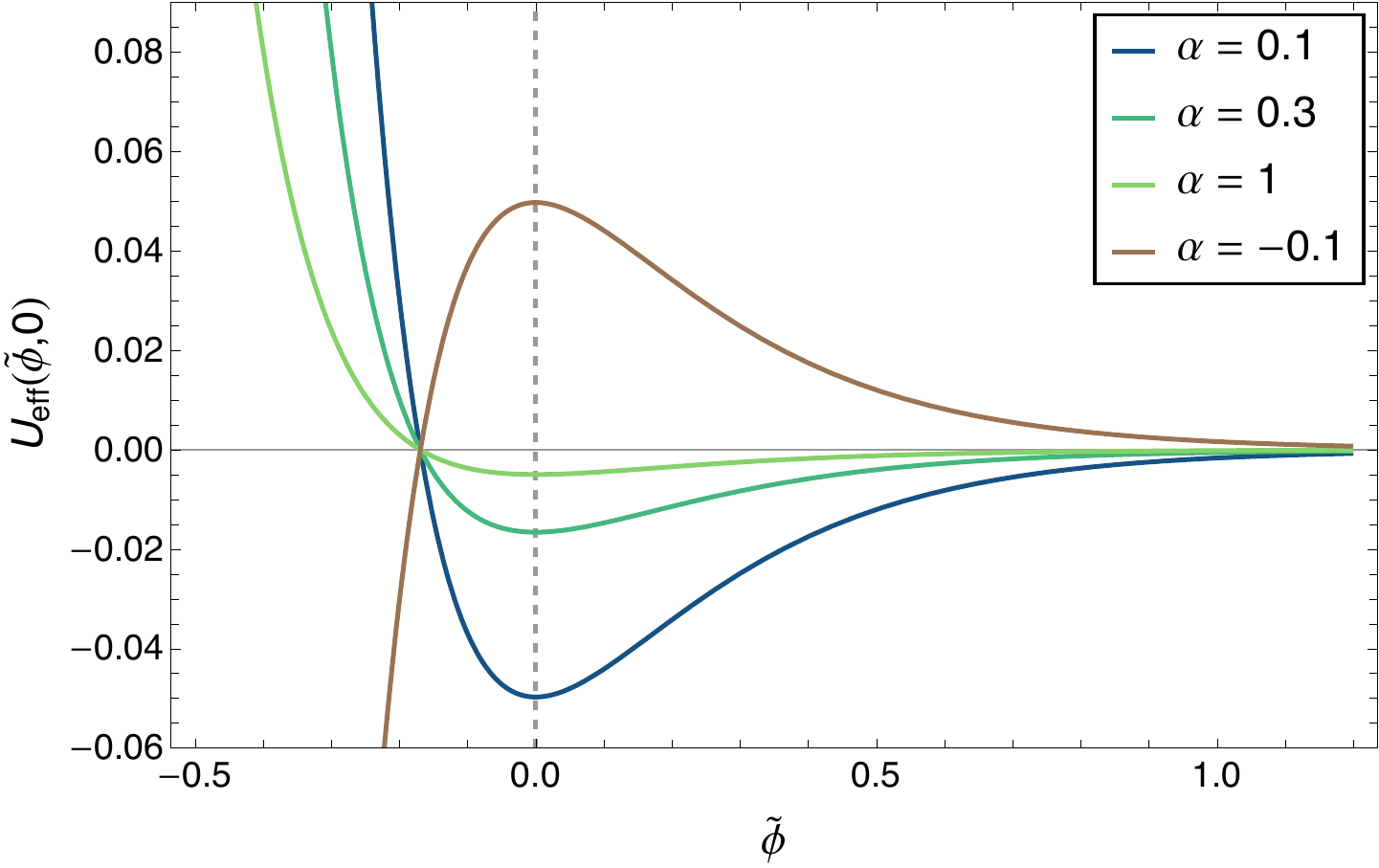}%
    \subcaption{Varying $\alpha$ with $T=0$.}%
    \label{fig:chame_pot_EI_a}%
  \end{minipage}%
  \begin{minipage}[t]{0.5\linewidth}%
    \centering%
    \includegraphics[keepaspectratio, width=0.9\linewidth]{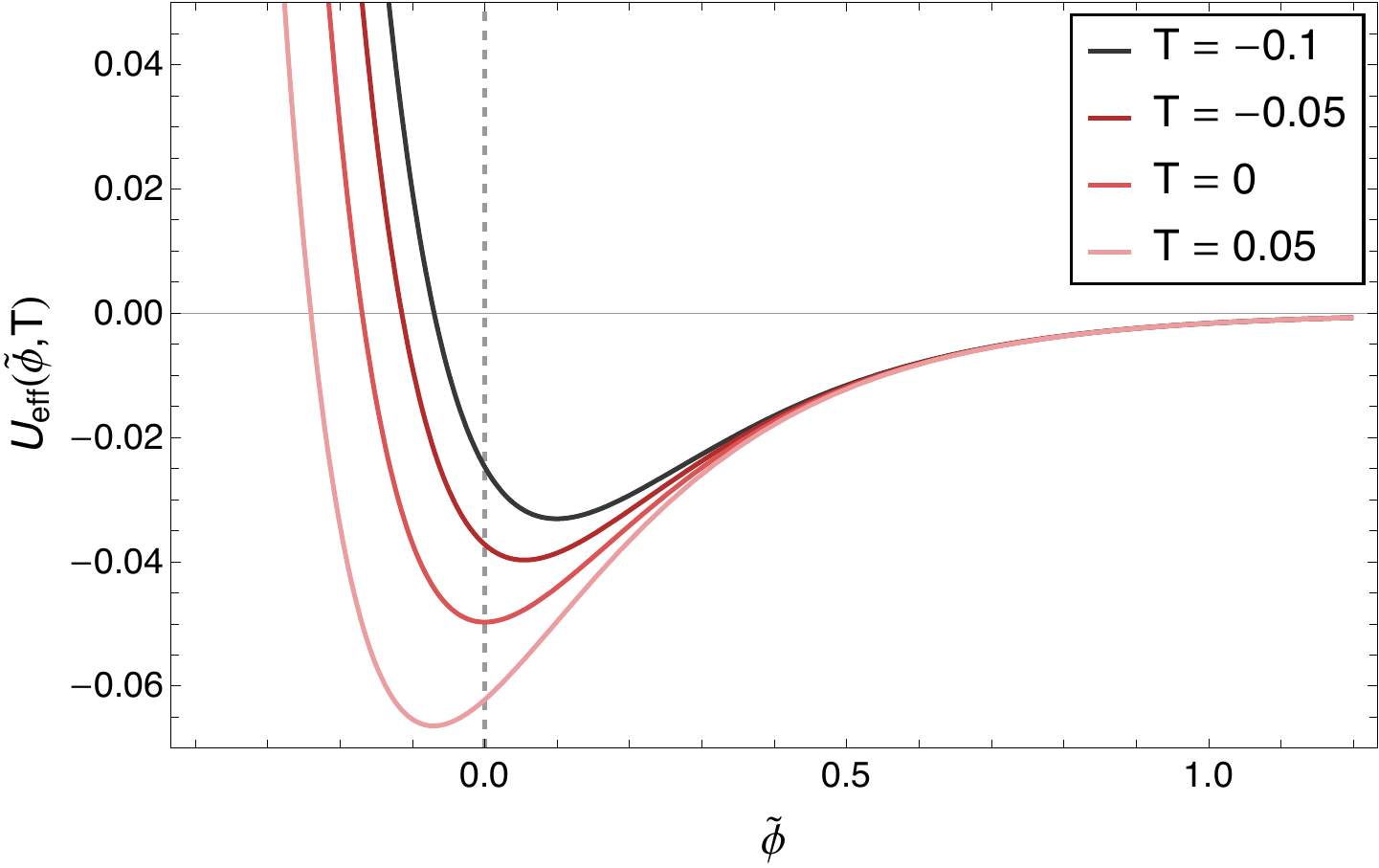}
    \subcaption{Varying $T$ with $\alpha=0.1$.}%
    \label{fig:chame_pot_EI_T}%
  \end{minipage}%
  \caption{
The chameleon potential \eqref{eq:chame_pot_Einstein} in the Einstein frame under the $R^2$ gravity \eqref{eq:R^2}. 
All quantities are dimensionless by $r_g = GM_{\odot}$.
$T$ is defined in the Jordan frame, and the bottom of the potential becomes shallower as $\alpha$ increases also in this frame. 
The potential minimum moves horizontally and becomes deeper as $T$ increases.
}
  \label{fig:chame_pot_EI}
\end{figure}%

For the $R^2$ gravity, the chameleon potential in the Einstein frame is derived as
\begin{align}
    U_{\mathrm{eff}} (\Tilde{\Phi}, T)
    = \frac{1}{8\kappa^2\alpha} \e^{-2\sqrt{\frac{2}{3}}\kappa\Tilde{\Phi}}
    \qty(2\, \e^{\sqrt{\frac{2}{3}}\kappa\Tilde{\Phi}}-1+2\kappa^2 \alpha T)\, ,
    \label{eq:chame_pot_Einstein}
\end{align}
up to a constant term. 
The plot for several $\alpha$ and $T$ are shown as \figref{fig:chame_pot_EI}. 
The chameleon mass is
\begin{align}
    m_{\Tilde{\Phi}}^2 
    = \frac{1}{6\alpha \qty(1-2\kappa^2 \alpha T)} \, .
\end{align}
The non-tachyonic condition of the scalaron field leads to the constraint on the energy-momentum in the Jordan frame:
\begin{align}
    T < \frac{1}{2 \kappa^2 \alpha} \, .
    \label{eq:cond_chamemass}
\end{align}
This constraint gives the upper limit on the possible value of central pressure (or energy density).

\bibliographystyle{apsrev4-1}
\bibliography{reference}

\end{document}